\documentclass[fleqn,usenatbib]{mnras}

\usepackage{newtxtext,newtxmath}

\usepackage[T1]{fontenc}
\usepackage{ae,aecompl}

\usepackage{graphicx}	            
\usepackage{amsmath}	            
\usepackage{amssymb}	            
\usepackage{cleveref}               
\usepackage[dvipsnames]{xcolor}     

\usepackage{bm}                     
\usepackage{siunitx}
\usepackage{xfrac}
\usepackage{ifthen}
\usepackage{booktabs}
\usepackage{xparse}
\usepackage[inline,shortlabels]{enumitem}
\usepackage[mathlines]{lineno}

\usepackage{eso-pic}

\AddToShipoutPictureBG*{%
  \AtPageUpperLeft{%
    \hspace{0.75\paperwidth}%
    \raisebox{-3.5\baselineskip}{%
      \makebox[0pt][l]{\textnormal{DES-2020-0563}}
}}}%

\AddToShipoutPictureBG*{%
  \AtPageUpperLeft{%
    \hspace{0.75\paperwidth}%
    \raisebox{-4.5\baselineskip}{%
      \makebox[0pt][l]{\textnormal{FERMILAB-PUB-20-480-AE}}
}}}%

\graphicspath{{include/}}

\newcommand{\redmagic}{redMaGiC}
\newcommand{\mat}[1]{\ensuremath{\bm{#1}}}

\newcommand{\slashfrac}[2]{\ensuremath{\left. #1 \middle/ #2 \right.}}
\DeclareMathOperator{\mad}{\mathrm{MAD}}
\DeclareMathOperator{\med}{med}

\DeclareMathOperator{\var}{Var}
\DeclareMathOperator{\diag}{diag}
\newcommand{\transpose}{\ensuremath{\top}}
\DeclareSIUnit{\sqdeg}{\text{square degrees}}
\DeclareSIUnit{\amin}{\text{arcmin}}
\makeatletter
\newcommand{\vvec}[1]{\ensuremath{\vec{#1}%
    \@ifnextchar^{\,}{}}%
}
\newcommand{\prob}{\@ifstar\probs\probn}
\newcommand{\cprob}{\@ifstar\cprobs\cprobn}
\makeatother
\newcommand{\probs}[1]{\ensuremath{P\!\left(#1\right)}}
\newcommand{\probn}[1]{\ensuremath{P(#1)}}
\newcommand{\cprobs}[2]{\ensuremath{P\!\left(#1\;\middle\vert#2\right)}}
\newcommand{\cprobn}[2]{\ensuremath{P(#1\;\vert\;#2)}}
\newcommand{\Romannumeral}[1]{\MakeUppercase{\romannumeral #1}}
\NewDocumentCommand{\mydefcitealias}{ m m }{%
    \defcitealias{#1}{#2}%
    \citeauthor{#1} (\citeyear{#1}; hereafter \citetalias{#1})%
}
\newcommand{\Nside}{N_{\rm side}}

\newcommand{\tct}{3$\times$2pt}
\newcommand{\mockchisq}{22.33}
\newcommand{\lamfac}{1.24}

\crefname{figure}{Fig.}{Figs.}
\crefname{equation}{Eqn.}{Eqns.}
\Crefname{figure}{Fig.}{Figs.}
\Crefname{equation}{Eqn.}{Eqns.}

\title[DES LSS Linear Systematics]{Linear Systematics Mitigation in Galaxy Clustering in the Dark Energy Survey Year 1 Data}

\author[E. L. Wagoner et al.]{
Erika L. Wagoner,$^{1}$\thanks{E-mail: wagoner47@email.arizona.edu (ELW)}
Eduardo Rozo,$^{1}$\thanks{E-mail: erozo@email.arizona.edu (ER)}
Xiao Fang,$^{2}$
\newauthor
Mart\'{i}n Crocce,$^{3}$
Jack Elvin-Poole,$^{4,5}$
and Noah Weaverdyck$^{6,7}$
\newauthor
(DES Collaboration)
\\
$^{1}$Department of Physics, University of Arizona, 1118 E. Fourth Street, Tucson, AZ, 85721, USA\\
$^{2}$Department of Astronomy and Steward Observatory, University of Arizona, 933 N Cherry Ave, Tucson, AZ, 85719, USA\\
$^{3}$Institut de Ci\'{e}nces de l’Espai, IEEC-CSIC, Campus UAB, Carrer de Can Magrans, s/n, 08193 Bellaterra, Barcelona, Spain\\
$^{4}$Center for Cosmology and Astro-Particle Physics, The Ohio State University, Columbus, OH, 43210, USA\\
$^{5}$Department of Physics, The Ohio State University, Columbus, OH, 43210, USA\\
$^{6}$Department of Physics, University of Michigan, 450 Church St, Ann Arbor, MI, 48109-1040, USA\\
$^{7}$Leinweber Center for Theoretical Physics, University of Michigan, 450 Church St, Ann Arbor, MI, 48109-1040, USA
}

\date{Accepted XXX. Received YYY; in original form ZZZ}

\pubyear{2020}

\begin{document}
\label{firstpage}
\pagerange{\pageref{firstpage}--\pageref{lastpage}}
\maketitle

\begin{abstract}
We implement a linear model for mitigating the effect of observing conditions and other sources of contamination in galaxy clustering analyses. Our treatment improves upon the fiducial systematics treatment of the Dark Energy Survey (DES) Year 1 (Y1) cosmology analysis in four crucial ways. Specifically, our treatment 
\begin{enumerate*}[1), itemjoin={{; }}, after={. }]
    \item does not require decisions as to which observable systematics are significant and which are not, allowing for the possibility of multiple maps adding coherently to give rise to significant bias even if no single map leads to a significant bias by itself
    \item characterizes both the statistical and systematic uncertainty in our mitigation procedure, allowing us to propagate said uncertainties into the reported cosmological constraints
    \item explicitly exploits the full spatial structure of the galaxy density field to differentiate between cosmology-sourced and systematics-sourced fluctuations within the galaxy density field
    \item is fully automated, and can therefore be trivially applied to any data set
\end{enumerate*}
The updated correlation function for the DES Y1 \redmagic{} catalog minimally impacts the cosmological posteriors from that analysis. Encouragingly, our analysis does improve the goodness of fit statistic of the DES Y1 3$\times$2pt data set ($\Delta \chi^2 = -6.5$ with no additional parameters). This improvement is due in nearly equal parts to both the change in the correlation function and the added statistical and systematic uncertainties associated with our method. We expect the difference in mitigation techniques to become more important in future work as the size of cosmological data sets grows.
\end{abstract}




\section{Introduction}
\label{sec:intro}

Galaxies are an important tool for studying the distribution of matter in the Universe and testing cosmological models. The correlation between the shapes of galaxies can be used to measure the weak gravitational lensing field, which is a direct measure of the total mass along lines of sight \citep[see, e.g.,][]{
1999ARA&A..37..127M,2003ARA&A..41..645R}. 
Because galaxies are a biased tracer of the mass, their clustering also carries within it cosmological information \citep[e.g.,][]{
2020arXiv200308277N}. 
In addition, the tangential distortion of background galaxies around the position of foreground galaxies---usually referred to as galaxy--galaxy lensing---can be used to study the correlation between the foreground galaxies and the matter around them \citep[e.g.,][]{
2004AJ....127.2544S,2006MNRAS.368..715M,2012PhRvD..86h3504Y}. 
The combination of all three measurements, sometimes referred to as ``\tct'' for the use of three 2-point functions, can break the degeneracy between the galaxy bias and the clustering amplitude of matter \citep[see, e.g.,][and references therein]{
2017MNRAS.470.2100K}. These \tct{} analyses are the focus of ongoing surveys such as the Kilo Degree Survey (KiDS; \citealt{2017MNRAS.465.1454H}), the Dark Energy Survey (DES; \citealt{2005astro.ph.10346T,2016MNRAS.460.1270D}), and the Hyper Suprime-Camera Survey (HSC; \citealt{2018PASJ...70S...4A}). Many other analyses have also applied a similar combined-probes approach to a variety of data sets \citep[see, e.g.,][]{2013MNRAS.432.1544M,2015ApJ...806....2M,2018MNRAS.476.4662V,2018MNRAS.474.4894J,2018PhRvD..98d3526A}. These types of combined analyses can also help to mitigate systematics that impact only one of the three 2-point measurements. 

Wide field stage \Romannumeral{2} dark energy experiments (such as the Sloan Digital Sky Survey [SDSS; \citealt{2000AJ....120.1579Y}], the WiggleZ Dark Energy Survey \citep{2010MNRAS.401.1429D} and the Canada-France-Hawaii Telescope Legacy Survey [CFHTLS; \citealt{2012SPIE.8448E..0MC}]) and stage \Romannumeral{3} dark energy experiments (e.g., KiDS, DES, HSC, and eBOSS) have provided imaging and spectra for hundreds of millions of galaxies, and stage \Romannumeral{4} experiments such as the Dark Energy Spectroscopic Instrument (DESI; \citealt{2013arXiv1308.0847L}), the Rubin Observatory's Legacy Survey of Space and Time (LSST; \citealt{2019ApJ...873..111I}), the \textit{Nancy Grace Roman Space Telescope} \citep{2015arXiv150303757S}, and \textit{Euclid} \citep{2011arXiv1110.3193L} are expected to increase that number substantially. As the number of observed galaxies increases, the statistical uncertainty on measurements made with them decreases. 
Consequently, our understanding and treatment of the systematic effects that impact galaxy clustering measurements must be improved if the uncertainties on the inferred cosmological parameters from such galaxy surveys are to remain statistics-dominated. 

There are a large number of potential contaminants that can result in this type of coherent fluctuations, e.g. star-galaxy separation, stellar occultation, extinction, and variations in observing conditions like airmass or sky brightness. Differentiating between the true cosmologically-sourced fluctuations and those caused by such survey properties has been the subject of many studies over the years \citep[see, e.g.,][and references therein]{2016MNRAS.457..786S,2018PhRvD..98d2006E,2020JCAP...03..044N,2020MNRAS.495.1613R,2020arXiv200714499W}
\citet{2020MNRAS.495.1613R} identify three broad categories of mitigation techniques:
\begin{enumerate*}[(a)]
    \item Monte Carlo simulation of fake objects;
    \item mode projection; and
    \item regression.
\end{enumerate*}

The first of these methods, involving injecting artificial sources into real images, is extremely promising. It results in forward-modeling the survey selection mask imposed by real imaging properties. Examples of this method include \citet{2013A&C.....1...23B} and \citet{2016MNRAS.457..786S}. However, this technique is computationally expensive, and therefore less utilized than the other methods.

Techniques utilizing mode projection typically involve down-weighting the spatial modes that are strongly correlated with survey properties by assigning a large variance to them. This technique has been explained and utilized in, e.g., \citet{1992ApJ...398..169R,2020JCAP...03..044N}.  
The variance of the estimated clustering increases as more survey properties are considered unless a threshold is used to limit the number of survey property maps. However, using such a threshold has been shown to introduce a bias in the resulting two-point function \citep{2016MNRAS.456.2095E}.

Regression-based techniques attempt to model the impact of the survey properties on the galaxy density, fitting the parameters of the model by cross-correlating the galaxies and systematic fluctuations or by using a least-squares estimate. For instance, \citet{2011MNRAS.417.1350R,2012ApJ...761...14H} fit for the impact of observing conditions in the correlation function and power spectrum, respectively. The disadvantage with this method is that any spurious correlation between the 2-point function of the galaxies and the survey properties will result in a correction, even if the fluctuations are not spatially related. This makes it easy to over-correct for systematic fluctuations which may bias the resulting correlation function estimate, although \citet{2020arXiv200714499W} show how the pseudo-Cl implementation of mode projection can be interpreted as an ordinary least squares regression approach that accounts for this over-correction.

As part of the analysis of the DES Y1 ``Gold'' data release, \mydefcitealias{2018PhRvD..98d2006E}{Paper~\Romannumeral{1}} also fit for the impact of survey properties, but using one of the alternative suggestions from \citet{2011MNRAS.417.1350R} of applying the corrections one at a time in order to account for potential correlations between different sources of systematic fluctuations. Briefly, the method of \citetalias{2018PhRvD..98d2006E} is as follows: the average number of galaxies per pixel $N_{\rm gal}$ is measured for all pixels with a survey property value $s$ within a bin $s \in [s_{\rm min}, s_{\rm max}]$ for one of the survey property maps, relative to the average number of galaxies per pixel in all pixels $\langle N_{\rm gal}\rangle$. A model is fit across all bins of the survey property values, and the $\Delta \chi^2$ for this model compared to a null test where $N_{\rm gal} / \langle N_{\rm gal}\rangle = 1$ is calculated. The significance of the survey property map is defined by comparing this $\Delta \chi^2$ to the sixty-eighth percentile of the equivalent quantity measured in \num{1000} contaminated Gaussian mock catalogs. This procedure is repeated for each survey property map and the maps are ranked by significance. A correction is applied for the most significant map to the measurements of $N_{\rm gal} / \langle N_{\rm gal}\rangle$, and the significance of each map is re-calculated. To avoid over-correction, this iterative process continues until none of the survey property maps have a significance above some target threshold. However, it is not necessarily the case that the effects of the various survey properties can be separated in this manner. For instance, this method precludes the possibility that significant systematic fluctuations can arise from the coherent contribution of multiple sources of systematics despite each individual survey property map being negligible by itself. Also, the analysis in \citetalias{2018PhRvD..98d2006E} included the spatial structure of the galaxy distribution only through the covariance in the galaxy densities binned by survey property. The analysis method introduced in this paper explicitly incorporates the density and spatial separations of neighboring pixels for determining the coefficients of the fluctuations sourced by survey properties: it is a much finer-grained look at that spatial structure.

Several other recent studies have attempted to use the regression-based technique directly with the galaxy density field while incorporating the spatial structure of the galaxy density field \citep[see, e.g.,][]{2016ApJS..224...34P,2017MNRAS.465.1831D}. However, as discussed in \citet{2020MNRAS.495.1613R}, these models are also often vulnerable to over-correction. The regression method used by \citet{2020MNRAS.495.1613R} differs from previous regression-based techniques in that it does not assume a functional form for the impact of the survey properties on the observed galaxy density. Instead, \citet{2020MNRAS.495.1613R} rely on a neural network approach and feature selection to achieve accurate systematic corrections without over-correction. However, this method fails to propagate the statistical and systematic uncertainties due to the correction into the error budget of the galaxy clustering signal.

In this paper, we implement an improved version of the linear model described in \citet{2016ApJS..224...34P}. Relative to that work, we reduce the number of free parameters by one by enforcing the condition that in the absence of systematic fluctuations, the observed galaxy density field will be equal to the true galaxy density field with a mean of zero (i.e., we do not include the constant term in equations 13 and 14 of that paper as a free parameter in our model). Our analysis explicitly incorporates the spatial clustering signal of the galaxy density field in an iterative approach, and mock catalogs are used to calibrate and correct for the residual bias due to over-correction. The combination of using a Markov chain Monte Carlo (MCMC) to fit our model and utilizing mock catalogs to correct for the bias allows us to estimate both the statistical and systematic uncertainty of our systematics-corrected galaxy correlation function. Our procedure therefore correctly inflates the error budget associated with the measurement of the galaxy correlation function, enabling us to trivially propagate these uncertainties into cosmological constraints downstream. We apply our model to the DES Y1 Gold \redmagic{} catalog, and compare our results to those from \citetalias{2018PhRvD..98d2006E} and \citet{2018PhRvD..98d3526A}. 

The paper is organised as follows: in \cref{sec:data}, we describe the \redmagic{} catalog and the survey properties we use. We describe our method in \cref{sec:method}. The generation of our mock catalogs and the results of the validation in the mocks is discussed in \cref{sec:mocks}. We determine the impact of our systematics correction on the uncertainty in the correlation function in \cref{sec:noise}. Our results are presented in \cref{sec:results}, and we summarize our findings in \cref{sec:conclusions}.

\section{Data}
\label{sec:data}

\begin{figure}
    \centering
    \includegraphics[width=\columnwidth,keepaspectratio]{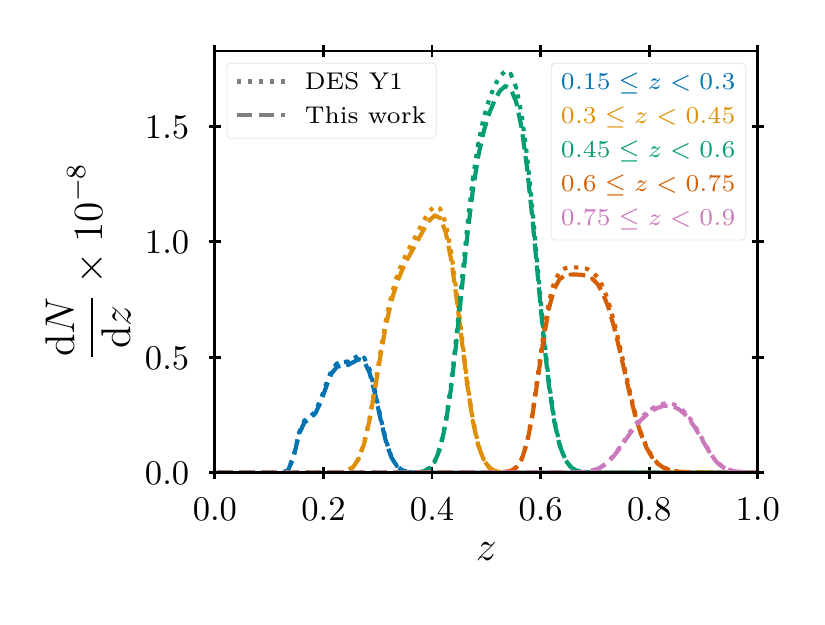}
    \caption{\label{fig:nz} The redshift distribution for each redshift bin, found by stacking Gaussian distributions with mean and standard deviation equal to the \redmagic{} redshift and error. The dashed lines are the distributions with our new mask, and the dotted lines of the corresponding colors are the corresponding distributions from \citetalias{2018PhRvD..98d2006E} in the same bins. The curves are not normalized, so differences in height are from the number of galaxies in the bin.}
\end{figure}

We will estimate and correct for systematic-sourced fluctuations in the density of the DES Year 1 \redmagic{} galaxy sample \citepalias{2018PhRvD..98d2006E}. We use the same redshift binning as the Y1 analysis, shown here in \cref{tab:zbins}, along with the number count and galaxy density in each bin. As described in \cref{sec:method}, our analysis leads us to remove survey regions with large systematic-sourced fluctuations. This cut removes \SI{\sim 3.5}{\percent} of the fiducial Y1 \redmagic{} footprint, for a final area of \SI{\approx 1274}{\sqdeg}. The counts and galaxy density after our systematic cut is shown in the fourth and fifth columns of \cref{tab:zbins}. \Cref{fig:nz} compares the redshift distributions in each bin before and after the systematics cuts. The dotted lines of various colors are the distributions for the full \redmagic{} sample, while the dashed lines of the same color are the distributions in the same bin after cutting based on systematics. The distributions are not normalized, so differences in height are caused by the difference in the number of galaxies before and after the cut.

\begin{table}
\centering
\begin{tabular}{lcccc}
\toprule
$z$ range & Y1 $N_{\rm gal}$ & Y1 $n_{\rm gal}$ & $N_{\rm gal}$ & $n_{\rm gal}$ \\
 &  & (\si{\per\square\amin}) &  & (\si{\per\square\amin}) \\ \midrule
$0.15 < z < 0.3$ & 63719 & 0.0134 & 61621 & 0.0134 \\
$0.3 < z < 0.45$ & 163446 & 0.0344 & 157800 & 0.0344 \\
$0.45 < z < 0.6$ & 240727 & 0.0506 & 231649 & 0.0505 \\
$0.6 < z < 0.75$ & 143524 & 0.0302 & 138450 & 0.0302 \\
$0.75 < z < 0.9$ & 42275 & 0.0089 & 40812 & 0.0089 \\ \bottomrule
\end{tabular}
\caption{\label{tab:zbins} The redshift binning with information about the number of galaxies and number density both from the DES Y1 analysis and the current analysis. The second and fourth columns are the total number of galaxies in each of the redshift bins, while the third and fifth give the galaxy density per square arcminute. Note that there is a change in the mask in going from the Y1 counts and number density of columns two and three to our own in columns four and five, which reduces the area by \SI{\sim 3.5}{\percent}.}
\end{table}

We estimate the galaxy correlation function using the \citet{1993ApJ...412...64L} estimator,
\begin{linenomath}\begin{equation}
    \label{eq:cfEstimator}
    \hat{w}(\theta) = \frac{DD - 2DR + RR}{RR},
\end{equation}\end{linenomath}
where $DD$, $DR$, and $RR$ are the number of pairs of galaxies with angular separation $\theta$ given a galaxy sample $D$ and a random catalog $R$. We measure the number of pairs using \texttt{TreeCorr}\footnote{\url{http://ascl.net/1508.007}} \citep{2004MNRAS.352..338J}, and our random catalog is the same one used by \citetalias{2018PhRvD..98d2006E}, except for the fact that we remove the random points in the survey regions excluded by our analysis.

We consider a total of 18 potential sources of systematics for the observed galaxy correlation function. Each of these is represented as a map which is pixelated on the sky using the \texttt{HEALPix}\footnote{\url{https://healpix.sourceforge.net}} \citep{2005ApJ...622..759G} pixelization scheme. The majority of the maps we consider are imaging properties from the DES Y1 `GOLD' catalog release \citep{2018ApJS..235...33D}. In each of the four bands ($griz$), we have maps of
\begin{enumerate}[(i),labelindent=\parindent,leftmargin=2\parindent]
    \item total exposure time;
    \item mean PSF FWHM;
    \item mean sky brightness, due to e.g. the moon; and
    \item mean airmass,
\end{enumerate}
 For all mean quantities, the value on a pixel in a given band is computed as the weighted mean over all exposures in that band which contribute to the pixel. The exposure time is instead the sum of the exposure times for each exposure contributing to the pixel. Unlike \citetalias{2018PhRvD..98d2006E}, we do not include any depth maps, as these depend on the other imaging properties in a complicated way, and therefore are not linearly independent from the other imaging properties---including the depth maps would be double counting the other imaging properties and would likely increase any over-correction biases that might exist. We therefore have \num{16} imaging property maps. We also consider contamination due to foreground stars, for which we use the stellar density map described in section~5 of \citetalias{2018PhRvD..98d2006E}. Galactic extinction is included using the dust opacity map from the Planck Collaboration \citep{2014A&A...571A..11P}. Both stellar density and extinction were considered in \citetalias{2018PhRvD..98d2006E}, but were not found to have significant correlation with the galaxy density and thus were ultimately excluded from that correction. We include both here because we do not want to preclude the possibility that they could still add coherently with other potential sources of contamination and thus impact the observed galaxy density. We note that the Planck dust map is known to have some positive correlation with the galaxy density in some redshift bins, as do most other Galactic dust maps \citep[see, e.g.,][]{2019ApJ...870..120C}. Collectively, we refer to the set of \num{18} imaging property, stellar density, and Galactic extinction maps as ``survey property maps'' or SPs. Where necessary, we use the routines of \texttt{healpy} \citep{2019JOSS....4.1298Z} for manipulating both survey property and density maps.

\section{Method}
\label{sec:method}

We determine the impact of observing conditions in the clustering of galaxies by relying on the spatial structure of the survey properties. Specifically, we estimate the extent to which the galaxy density maps are contaminated by systematic fluctuations by measuring the extent to which the galaxy density map traces the various survey property maps.

We begin by constructing a low-resolution ($\Nside=128$) map of the galaxy density field. This choice limits the number of empty pixels to a small percentage (\SI{\leq 10}{\percent}) of the total pixels. Working at this resolution, the average number of galaxies per pixel is \num{\geq 10} at all redshifts.

We degrade the resolution of our survey property maps to match the resolution of our galaxy density map, properly accounting for the masked portions of every pixel. Specifically, the degraded survey property map $\mathcal{S}^\prime$ is related to the original survey property map $\mathcal{S}$ via
%
\begin{linenomath}\begin{equation}
    \label{eq:degradedSys}
    \mathcal{S}^{\prime\, j} = \frac{\sum_{i \in j} \mathcal{S}^i f^i}{\sum_{i \in j} f^i},
\end{equation}\end{linenomath}
%
where $f^i \in [0, 1]$ is the fraction of pixel $i$ (at the original map resolution) that is detected in the footprint. The sums are over all high resolution pixels $i$ that fall within low resolution pixel $j$. We also degrade the pixel fraction map $f^i$, such that the fraction $f^{\prime j}$ of low resolution pixel $j$ in the footprint is related to the high resolution fraction by
%
\begin{linenomath}\begin{equation}
    \label{eq:degradedFrac}
    f^{\prime j} = \frac{1}{\tilde{N}} \sum_{i \in j} f^i,
\end{equation}\end{linenomath}
%
where $\tilde{N}$ is the number of high resolution pixels within a low resolution pixel.

The degraded survey property maps are transformed into standardized fluctuation maps as follows.  Let $\mathcal{S}^{\prime i}_\beta$ be the value of survey property map $\beta$ on low resolution pixel $j$.
We define the mean $\overline{\mathcal{S}}_\beta$ and fluctuation scale $\hat{\sigma}_\beta$ of $\mathcal{S}^{\prime j}_\beta$ via
%
\begin{linenomath}\begin{align}
    \overline{\mathcal{S}}_\beta &\equiv \slashfrac{\sum_{j = 1}^{N_{\rm pix}} f^{\prime j} \mathcal{S}^{\prime j}_\beta}{\sum_{j = 1}^{N_{\rm pix}} f^{\prime j}} \label{eq:sysMean} \\
    \intertext{and}
    \hat{\sigma}_\beta &\equiv 1.4826 \mad\!\left(\mathcal{S}^{\prime j}_\beta\right) \label{eq:sysSTD} \, ,
\end{align}\end{linenomath}
%
The median absolute deviation in \cref{eq:sysSTD} is
%
\begin{linenomath}\begin{equation*}
    \mad\!\left(\mathcal{S}^{\prime j}_\beta\right) \equiv \slashfrac{\sum_{j = 1}^{N_{\rm pix}} \left|\mathcal{S}^{\prime j}_\beta - \med\!\left(\mathcal{S}^{\prime j}_\beta\right)\right|}{N_{\rm mask}}\, ,
\end{equation*}\end{linenomath}
%
where $N_{\rm mask}$ is the number of pixels not removed by the mask. The ``fluctuation scale'' $\hat \sigma_\beta$ defined above is an estimator of the standard deviation for Gaussian fluctuations, but its value is more robust to outliers than estimates based on the sample variance. The standardized fluctuation map for survey property $\beta$ is defined as
%
\begin{linenomath}\begin{equation}
    \label{eq:standardSys}
    S^j_\beta \equiv \frac{\mathcal{S}^j_\beta - \overline{\mathcal{S}}_\beta}{\hat{\sigma}_\beta}\, .
\end{equation}\end{linenomath}
%

Rather than working with the fluctuation maps themselves, we construct an orthogonal map eigenbasis as follows. We assume the survey properties on each pixel are an independent random realization from an $N_{\rm maps}$-dimensional distribution. We find the covariance matrix $\mat{C}$ of the standardized maps at the fit resolution, where
%
\begin{linenomath}\begin{equation*}
    \mat{C}_{\alpha \beta} = \langle \left(S_\alpha - \langle S_\alpha \rangle \right) \left(S_\beta - \langle S_\beta \rangle \right) \rangle ,
\end{equation*}\end{linenomath}
%
and $\langle \cdot \rangle$ is the spatial average over all observed pixels. We define the rotation matrix $\mat{R}$ from the eigenvectors of $\mat{C}$ such that
\begin{linenomath}\begin{equation*}
    \mat{C} = \mat{R} \mat{D} \mat{R}^\transpose ,
\end{equation*}\end{linenomath}
where $\mat{D}$ is a diagonal matrix with the eigenvalues of $\mat{C}$ along the diagonal. The rotated and standardized survey property value for map $\alpha$ on pixel $j$ is
%
\begin{linenomath}\begin{equation}
    \label{eq:sAlpha}
    s^j_\alpha \equiv \mat{R}^\transpose_{\alpha \beta} S^j_\beta .
\end{equation}\end{linenomath}
%
Each $s^j_\alpha$ is, therefore, a linear combination of the fluctuations in the original SP maps $\{\mathcal{S}^{\prime\, j}_\beta\}$ on a given pixel. For the rest of the paper, unless otherwise noted, the term ``SP'' refers to the eigenmap $s^j_\alpha$ of \cref{eq:sAlpha} rather than the original survey property map $\mathcal{S}^i_\beta$. 

Since fluctuations in the density field can't be sensitive to a constant non-zero SP value---any non-zero constant would simply shift the mean value of the galaxy density field---, the observed galaxy density must only depend upon the fluctuations of the SPs. Thus, we write $\delta^j_{\rm obs} \equiv \delta_{\rm obs}\!\left(\{s_\alpha^j\}\right)$, where $\{s_\alpha^j\}$ is a vector containing the value of pixel $j$ across all SP maps $\alpha$. Expanding around $\{s_\alpha^j\} = \vvec{0}$ to first order, we have
%
\begin{linenomath}\begin{equation}
    \label{eq:deltaObs}
    \delta_{\rm obs}^j\!\left(\{s_\alpha^j\}\right) \approx \delta^j_{\rm true} + \sum_\alpha a_\alpha s^j_\alpha ,
\end{equation}\end{linenomath}
%
where the coefficient $a_\alpha$ is the derivative of $\delta_{\rm obs}$ with respect to $s_\alpha$ at $\{s_\alpha^j\} = \vvec{0}$. Note that any impact on the monopole of the galaxy density field by the survey properties gets absorbed into the mean observed galaxy density, and therefore has no impact on the galaxy fluctuations. Since our expansion is at first order, we can ignore the monopole as any impact with couplings to the linear perturbations would be second order. In the expansion, we have used the fact that $\delta^j_{\rm obs}\!\left(\{s_\alpha^j\} = \vvec{0}\right) = \delta^j_{\rm true}$, where $\delta^j_{\rm true}$ is the true galaxy overdensity on pixel $j$.  We have also assumed that the impact of SP on the galaxy density field is local: the SP in pixel $j$ only impact the galaxy density at pixel $j$.

Our task is to find the set of coefficients $\{a_\alpha\}$ in \cref{eq:deltaObs}. We do this by fitting the likelihood \cprob*{\vvec{\delta}_{\rm obs}}{\vvec{\delta}_{\rm sys}} of the observed overdensity map given the systematics map $\vvec\delta_{\rm sys} \equiv\sum_\alpha a_\alpha \vvec{s}_\alpha$, where the vector symbol denotes the full map. As discussed below, our procedure allows for covariance between pixels, so that this likelihood distribution does not in general reduce to a product over all pixels. We assume a Gaussian likelihood for $\vvec{\delta}_{\rm obs}$. This explains why it is important for the mean number of galaxies in the galaxy density map to be large. We test our sensitivity to using a Gaussian distribution in \cref{sec:mockResults}. The ensemble average over realizations of the observed density field at fixed systematics is simply
%
\begin{linenomath}\begin{equation}
    \label{eq:deltaObsMean}
    \left\langle \vvec{\delta}_{\rm obs}\right\rangle = \vvec{\delta}_{\rm sys} .
\end{equation}\end{linenomath}
%
We can thus write our Gaussian likelihood for $\vvec{\delta}_{\rm obs}$ as
%
\begin{linenomath}\begin{align}
    \ln \cprob*{\vvec{\delta}_{\rm obs}}{\vvec{\delta}_{\rm sys}} = &-\frac{1}{2} \log \left|\mat{\Sigma^{\rm obs}}\right| \nonumber \\ 
    &- \frac{1}{2} \left(\vvec{\delta}_{\rm obs} - \vvec{\delta}_{\rm sys}\right)^{\!\transpose} \left(\mat{\Sigma^{\rm obs}}\right)^{-1} \left(\vvec{\delta}_{\rm obs} - \vvec{\delta}_{\rm sys}\right) , \label{eq:like}
\end{align}\end{linenomath}
%
where we have dropped all constant terms, and again
\begin{equation}
    \vvec{\delta}_{\rm sys} = \sum_\alpha a_\alpha s^j_\alpha.
\end{equation}
The model parameters characterizing $\vvec{\delta}_{\rm sys}$ are the coefficients $a_\alpha$ for each survey property, which we aim to recover from the data. With this notation, both $\vvec{\delta}_{\rm sys}$ and $\vvec{\delta}_{\rm obs}$ are vectors of length $N_{\rm pix}$ and $\mat{\Sigma^{\rm obs}}$ is an $N_{\rm pix} \times N_{\rm pix}$ matrix, where $N_{\rm pix}$ is the number of pixels within the footprint (i.e. the number of observed pixels).

The covariance matrix for our likelihood can be written as the sum of two terms,
%
\begin{linenomath}\begin{equation}
    \label{eq:deltaObsCov}
    \mat{\Sigma^{\rm obs}} = \mat{\Sigma^{PN}} + \mat{\Sigma^{SV}} .
\end{equation}\end{linenomath}
%
The first term contains the Poisson noise in the density field, and takes the form
%
\begin{linenomath}\begin{equation*}
    \mat{\Sigma^{PN}}_{jk} = \sigma_g^2 \delta_{jk} ,
\end{equation*}\end{linenomath}
%
where $\sigma_g$ is a constant for which we can fit and $\delta_{jk}$ is the Kronecker delta. It will become clear shortly why we allow $\sigma_g$ to be an unknown constant, rather than fixing it to the Poisson expectation. The second term in \cref{eq:deltaObsCov} accounts for the sample variance.

We fit for our SP coefficients in two iterations. During the first iteration, we assume there is no sample variance, so that $\mat{\Sigma^{\rm obs}}$ is diagonal. In this case, we can analytically solve for the variance $\sigma_g^2$ and coefficients $\{a_\alpha\}$ that minimize the likelihood in \cref{eq:like} by solving the simultaneous set of equations obtained when setting all of the partial derivatives with respect to the survey parameter coefficients and $\sigma_g^2$ to zero. We are also able to find the $19\times19$-dimensional parameter covariance matrix analytically as the inverse of the Hessian matrix evaluated at the minimum---we use this parameter covariance matrix (excluding the row and column corresponding to $\sigma_g^2$) in the second iteration to select random starting locations within the \num{18}-dimensional parameter space. 

Once we complete our first iteration, we use our results to estimate $\hat{\vvec{\delta}}_{\rm true}$. We then define $\mat{\Sigma^{SV}}$ via
%
\begin{linenomath}\begin{equation*}
    \mat{\Sigma^{SV}}_{jk} = (1 - \delta_{jk})\,  \hat{w}_{\rm true}\!\left(\theta_{jk}\right) ,
\end{equation*}\end{linenomath}
%
where $\hat{w}_{\rm true}$ is the correlation function of our estimated true overdensity field $\hat{\vvec{\delta}}_{\rm true}$ and $\theta_{jk}$ is the angular separation between pixels $j$ and $k$. We artificially set the diagonal elements of $\mat{\Sigma^{SV}}$ to zero because we cannot differentiate between the sample variance and Poisson noise within a single pixel. This also explains why we treated $\sigma_g$ as an unknown constant: $\sigma_{\rm g}$ is really the sum of the Poisson and zero-offset sample variance terms. We therefore continue to use the $\sigma_g$ obtained from the minimization in the first iteration as the only term on the diagonal of $\mat{\Sigma^{\rm obs}}$ in the second iteration.

We use the resulting ``Poisson'' and sample variance noise estimates to refit for the coefficients of each of the SP parameters. In the second iteration, we use a Markov Chain Monte Carlo (MCMC) algorithm (specifically \texttt{emcee}; \citealt{2013PASP..125..306F}) to sample our parameter space and estimate the posterior distribution while holding both $\mat{\Sigma^{PN}}$ and $\mat{\Sigma^{SV}}$ fixed. Our best fit coefficients after the second iteration are the mean parameter values from the chain\footnote{We run our chain with \num{36} walkers for \num{1000} steps each. We do not use a burn-in when fitting to the real data as we generate the initial positions by drawing from a multivariate Gaussian with a mean and covariance matrix given by the coefficients and parameter covariance from the first iteration. We use a burn-in of \num{300} steps per walker when fitting to mock catalogs.}. To check for convergence, we look at the shift in the coefficients between the first and second halves of each chain relative to the error from the chain. We find a median shift (over all \num{18} parameters) of \numlist[list-final-separator = {, and }]{0.19;0.29;0.18;0.26;0.14} for redshift bins \numrange[range-phrase = { through }]{1}{5} respectively, and the worst convergence in any single parameter for each redshift bin is \numlist[list-final-separator = {, and }]{0.60;0.72;0.55;0.53;0.34}. We have verified that using the coefficients from the second iteration to update $\mat{\Sigma^{SV}}$ and performing a second MCMC (i.e. getting a third iteration of the coefficients) does not have a significant impact on our results.

Once we have our coefficients, we correct for the effect of systematic fluctuations on the correlation function. We do so by defining weights for each galaxy based on the systematics map value on the pixel containing the galaxy. For calculating galaxy weights, we use the systematics map at a resolution of $\Nside = 4096$. While we must fit at low resolution to ensure that our likelihood is roughly Gaussian, the fundamental assumption of our method is that survey properties only produce local modulations of the galaxy density field. Since our model is linear, all the local modulations add together when smoothing to go to lower resolution, so the relation between the survey properties and the galaxy density must be the same at low and high resolution. We standardize and rotate the high resolution maps as we did with the low resolution maps, but we use the mean, fluctuation scale, and rotation matrix determined from the low resolution maps for the purposes of defining the high resolution eigen-maps. This is critical, as the definition of the maps must match that employed in our fits. The weight for a galaxy on high-resolution pixel $i$ is
%
\begin{linenomath}\begin{equation}
    \label{eq:weight}
    w^i = \frac{1}{1 + \sum_\alpha a_\alpha s^i_\alpha} .
\end{equation}\end{linenomath}
%
We refer to the correlation function measured using these weights as $w_{\rm corr0}$. As previously mentioned, when calculating the systematics-corrected correlation function, we also exclude any galaxies on pixels with $\delta_{\rm sys}^i > 0.2$. This should restrict us to only areas of the sky where our first order approximation is valid. The resulting footprint is \SI{\sim 3.5}{\percent} smaller than the original Y1 footprint, and a total of \num{23359} galaxies are removed across all redshift bins. We expect complications due to the interpolation to higher resolution to be small as we find that only \SI{\sim 12.8}{\percent} (\SI{\sim 0.8}{\percent}) of galaxies have a weight that differs from unity by more than \SI{10}{\percent} (\SI{20}{\percent}) before applying the cut based on $\delta_{\rm sys}^i$.

The above procedure tends to over-correct the data for the impact of SPs. We calibrate the amount of over-correction in the correlation function from our method using mock galaxy catalogs, and use these to de-bias our procedure, which will result in an updated systematics-corrected correlation function estimate $w_{\rm corr1}$. The details of this de-biasing are presented in the next section. We describe how we incorporate statistical and systematic uncertainties due to our correction in the error budget of the observed correlation function in \cref{sec:noise}.

\section{Methodology Validation with Mock Catalogs}
\label{sec:mocks}

There are three potential sources of systematic bias in our analysis. These are, in no particular order, 
\begin{enumerate*}[(i)]
\item the first order approximation from \cref{eq:deltaObs} is not accurate,
\item the Gaussian likelihood is not correct, and
\item the estimates of the SP coefficients are noisy and too much correlation is removed from the data, an effect usually referred to as over-correction.
\end{enumerate*}
As mentioned in \cref{sec:data}, we restrict our final data set to pixels where the linear prediction of the SP-sourced galaxy density fluctuations are \num{\leq 0.2}. This serves to minimize potential biases from non-linear responses in the systematics correction. We test the robustness of our methodology to non-Gaussian fields and noise by testing it on log-normal mock galaxy catalogs. We further use these catalogs to calibrate the bias in our method due to over-correction.

\subsection{Mock Catalog Generation}
\label{sec:mockData}

To create our log-normal mock catalogs, we use the fiducial cosmological parameters from \citetalias{2018PhRvD..98d2006E}: $\Omega_m = 0.295$, $A_s = 2.260574 \times 10^{-9}$, $\Omega_b = 0.0468$, $h = 0.6881$, and $n_s = 0.9676$. We run \texttt{CAMB} \citep{2000ApJ...538..473L,2012JCAP...04..027H} and \texttt{Halofit\_Takahashi} \citep{2003MNRAS.341.1311S,2012ApJ...761..152T} using \texttt{CosmoSIS} \citep{2015A&C....12...45Z} to compute the angular galaxy clustering power spectrum. We then use this power spectrum to generate a log-normal random field for the true galaxy over-density, $\delta_{\rm true}$, in each of our five redshift bins via the code \texttt{psydocl}\footnote{\url{https://bitbucket.org/niallm1/psydocl/src/master/}}. This galaxy density field is generated at high resolution ($\Nside=4096$). When appropriate (i.e. depending on the test being pursued, see below), we add systematic fluctuations to the galaxy density field using our linear model. We then calculate the expected number of galaxies in each pixel, taking into account the masked fraction in each pixel.  Finally, we randomly place $N$ galaxies within each pixel, where $N$ is a Poisson realization of the expected number of galaxies.

We generate \num{100} independent realizations of $\delta_{\rm true}$ for each redshift bin. Each realization is then used to create two mock catalogs, one with no SP contamination and another with SP applied using the best fit coefficients from our analysis of the DES Y1 data set. We refer to these as uncontaminated and contaminated mocks, respectively. Note that while both the uncontaminated and contaminated mocks share the same underlying over-density fields, they have different Poisson realizations.

We use our methodology from \cref{sec:method} to estimate the impact of SPs in our mock galaxy catalogs, and compare the resulting corrected correlation function to the underlying true mock galaxy correlation function. To increase computational efficiency, we restrict our mock catalogs to the final mask employed in our analysis of the DES Y1 galaxies. That is, we do not re-apply the $\delta_{\rm sys}^i \leq 0.2$ cut in every mock. Doing so would have forced us to recompute random pairs for every mock due to slight differences in the final footprint. Because systematic fluctuations are linear in the mock catalog by construction, this additional restriction has no bearing on the conclusions drawn from our simulations. Unfortunately, this also means our mock catalogs do not allow us to test how sensitive our method is to non-linear contamination.

We test whether our contaminated mock galaxy catalogs have comparable levels of SP contamination to the data as follows. For the data and both sets of mock galaxy catalogs we compute the raw observed correlation function, and the corrected correlation function $w_{\rm corr0}$ as described in \cref{sec:method}. We then calculate the difference between these two correlation functions in all three cases.

\begin{figure*}
    \centering
    \includegraphics[width=\textwidth,keepaspectratio]{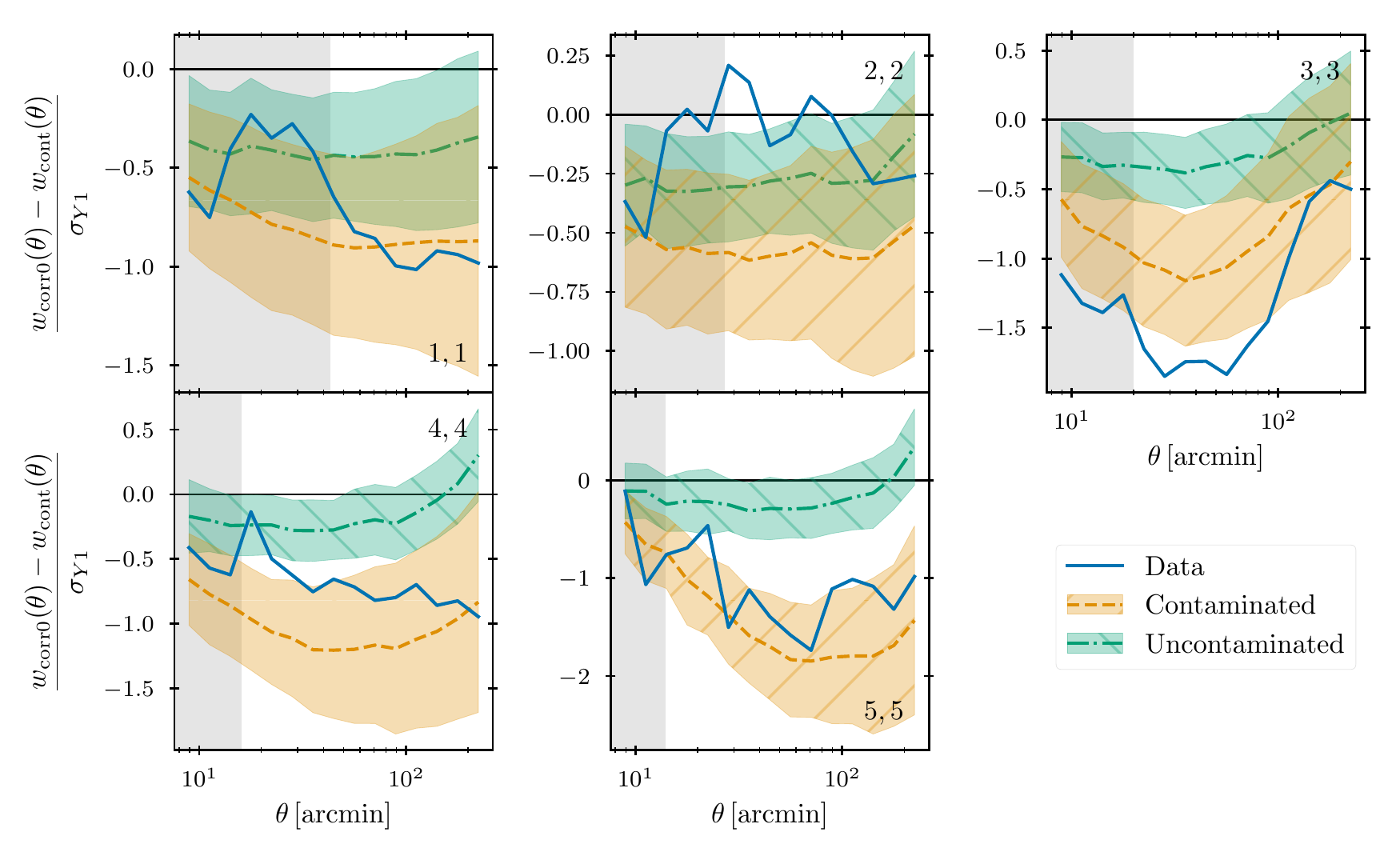}
    \caption{\label{fig:correctionSize} Comparison of the bias between the systematics-corrected ($w_{\rm corr0}$) and uncorrected ($w_{\rm cont}$) correlation functions for the DES Y1 data and the uncontaminated and contaminated mocks, relative to the DES Y1 errors (see text for details). The blue solid line is the result for the data. The mean and sample standard deviation for the contaminated mocks is shown as the orange dashed line and orange shaded region, while the green dash-dotted line and green shaded region show the same for the uncontaminated mocks. These error regions do not include the correction factor discussed in \cref{sec:mockResults}. By eye, we see $1 \sigma$ agreement between the contaminated mocks and the data for three of the five redshift bins, and $2 \sigma$ agreement in bins 2 and 3. The gray shaded region is once again the small scale cut used by \citetalias{2018PhRvD..98d2006E}.}
\end{figure*}

The blue solid line in \cref{fig:correctionSize} shows the biased systematic correction of the DES Y1 \redmagic{} data computed using the first iteration of our method, while the orange dashed line is the mean correction from the 100 contaminated mock galaxy catalogs. The green dashed-line is the mean of the uncontaminated galaxy catalogs. The width of the bands show the sample standard deviation for each of the two sets of mocks. It is immediately apparent that the amplitude of the systematic correction in our uncontaminated mocks is significantly smaller than that of the data in redshift bins 3, 4, and 5. That is to say, we have robustly detected the presence of systematic fluctuations in the DES Y1 data set. More generally, the correction derived from our contaminated mocks is comparable to that in the data, particularly for the redshift bins that exhibit strong systematic fluctuations. Thus, \cref{fig:correctionSize} provides evidence that the contaminated mock galaxy catalogs used in our analysis are a reasonable match to the data.

\subsection{Methodology Validation: Recovery of the SP Coefficients}
\label{sec:mockResults}

\begin{figure}
    \centering
    \includegraphics[width=\columnwidth,keepaspectratio]{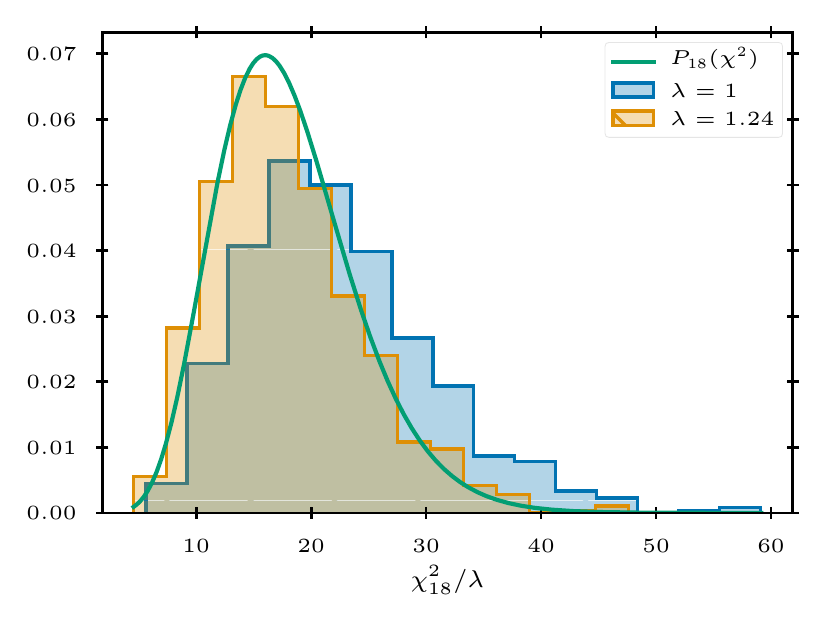}
    \caption{\label{fig:chi2hist} The distribution of $\chi^2$, as defined in \cref{eq:chi2}, for all contaminated and uncontaminated mocks in all redshift bins. The blue histogram is the original distribution. The orange histogram is the result of re-scaling every $\chi^2$ by $\sfrac{18}{\langle \chi^2\rangle}$. The green line is the expected $\chi^2$ distribution with \num{18} degrees of freedom, for reference. Note that both histograms are normalized.}
\end{figure}

We fit for the SP coefficients in both sets of \num{100} mocks for each redshift bin, for a total of \num{1000} independent mock catalogs to be analyzed. Because we know the SP coefficients used to generate the mocks, we can test whether we correctly recover the input coefficients with our analysis. To do so, we calculate the $\chi^2$ of the mean coefficients estimated from our posterior and the input for each mock. That is, for each mock catalog $\nu$ we compute
%
\begin{linenomath}\begin{equation}
    \label{eq:chi2}
    \chi^2_\nu = \left(\{\hat{a}_\alpha\}_\nu - \{a_\alpha\}_{0, \nu}\right)^\transpose \hat{\mat{C}}_\nu^{-1} \left(\{\hat{a}_\alpha\}_\nu - \{a_\alpha\}_{0, \nu}\right) ,
\end{equation}\end{linenomath}
%
where $\{a_\alpha\}_{0, \nu}$ is the input vector of \num{18} coefficients used in generating mock catalog $\nu$, $\{\hat{a}_\alpha\}_\nu$ is the mean vector of the posterior from our analysis for mock $\nu$ with length \num{18}, and $\hat{\mat{C}}_\nu$ is the parameter covariance matrix estimated from the MCMC chain for mock $\nu$ with dimensions $18\times18$. We show the distribution of the $\chi^2_\nu$ statistics for all \num{1000} mocks as the blue histogram in \cref{fig:chi2hist}. For reference, the green line is the expected $\chi^2$ distribution for \num{18} degrees of freedom, \num{18} being the number of SPs. It is clear that the distribution of $\chi^2$ values is biased relative to our expectation.

\citet{2007A&A...464..399H} pointed out that noise in the covariance matrix biases $\chi^2$ statistics. In our case, the noise in the covariance matrix is only partly due to a finite number of realizations in the MCMC: noise in the data will also generate noise in the empirically estimated covariance matrix, which will in turn bias the recovered $\chi^2$. In the absence of a first principles prescription for the expected bias in our analysis, we adopt an ad-hoc correction by demanding the average $\chi^2$ over all our simulations be equal to the number of degrees of freedom in the problem (\num{18}). That is, we de-bias every $\chi^2$ value by dividing it by the factor $\lambda \equiv \mockchisq/18=\lamfac$. The resulting distribution is shown as the orange histogram in \cref{fig:chi2hist}, which is now an excellent match to expectations.

As discussed in \citet{2007A&A...464..399H}, the bias due to noise in the covariance matrix estimate propagates into the parameter posteriors. Consequently, we increase the statistical uncertainty in our recovered corrections for the correlation function by a factor of $\sqrt{\lamfac}$. The fact that our recovered distribution of $\chi^2$ values matches expectation implies that we are successfully recovering the input systematic coefficients within our re-scaled noise estimate.

\subsection{Over-correction Calibration}
\label{sec:calibration}

\begin{figure*}
    \centering
    \includegraphics[width=\textwidth,keepaspectratio]{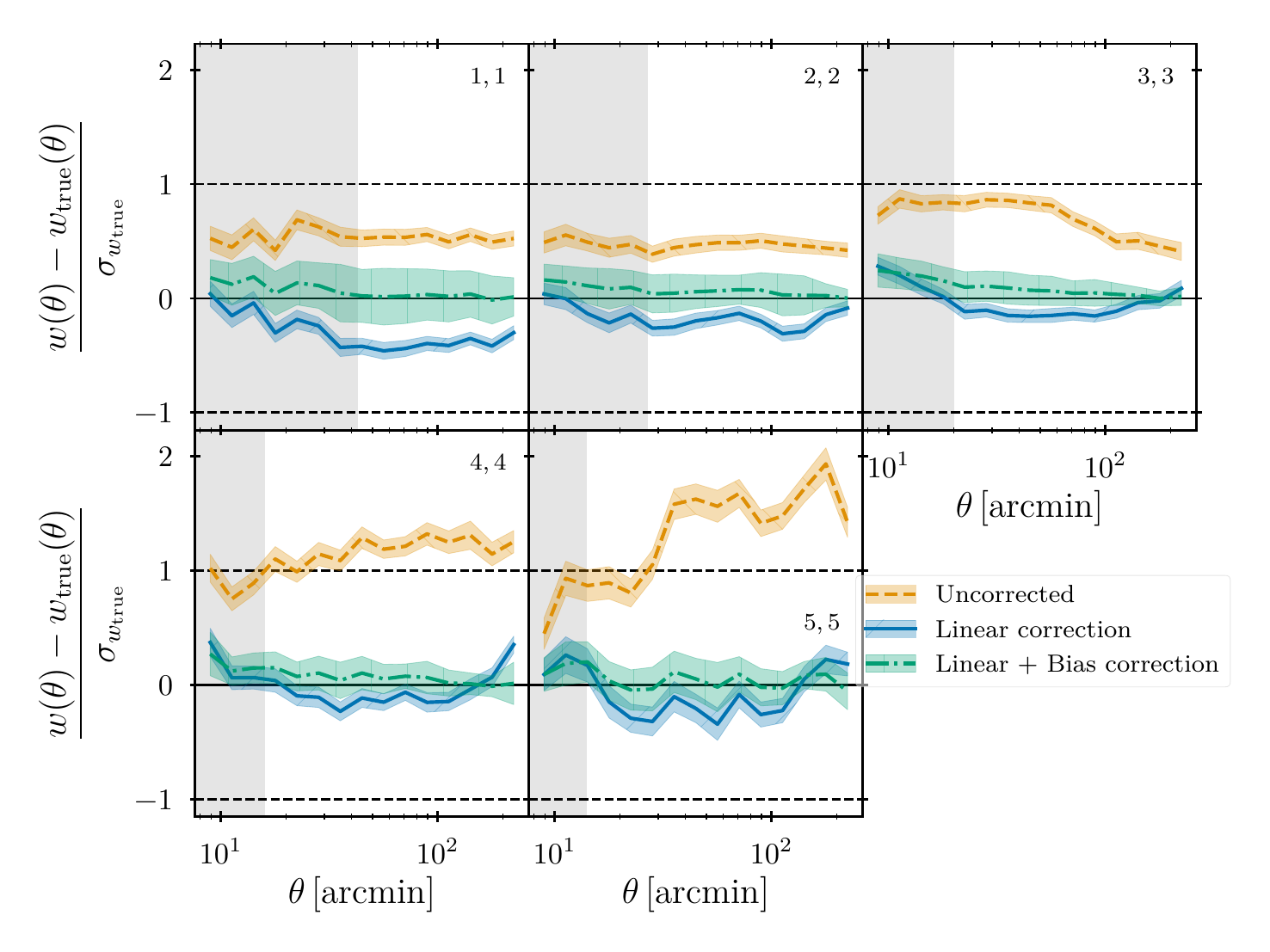}
    \caption{\label{fig:corrections} The difference between the various correlation functions for the contaminated mocks and the true correlation function. The orange dashed line shows the offset for the correlation function without any corrections. The blue solid line shows the offset when the systematics weights are applied, but no bias correction is used. The green dash-dotted line is the final offset, with both the systematics weights and the bias correction. Each line is the mean for the \num{100} mocks, and the shaded regions are the error on the mean. Note that the offset is also divided by the sample standard deviation of the true correlation function. We only scales with $\theta > \SI{8}{\arcmin}$ for clarity. The gray shaded region shows the small scale cut used by \citetalias{2018PhRvD..98d2006E}, so any scales within that region will not impact the cosmology results.}
\end{figure*}

The orange dashed line and shaded band in \cref{fig:corrections} show the mean and $1 \sigma$ region for the difference between the observed and true correlation functions of our \num{100} independent systematics-contaminated mock catalogs, in units of the statistical uncertainty of the DES Y1 analysis. The $1 \sigma$ region is computed as the error on the mean. The blue solid line and shaded band are the same as the orange, but for the systematics-corrected correlation function with no bias correction (i.e. $w_{\rm corr0}$). While there is a significant improvement when going from no correction to our systematics correction, it is also clear that our method somewhat over-corrects the data.

\begin{figure*}
    \centering
    \includegraphics[width=\textwidth,keepaspectratio]{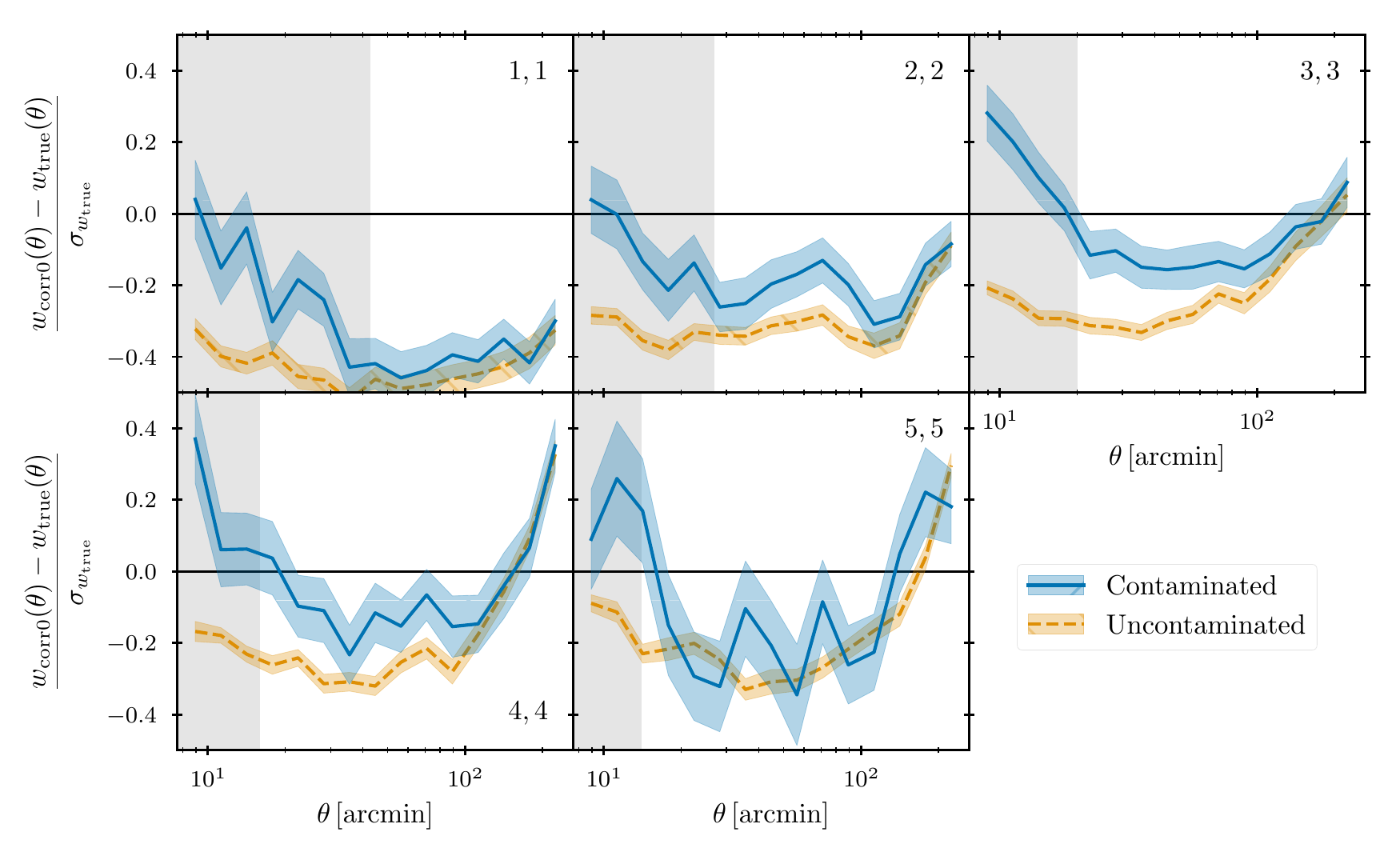}
    \caption{\label{fig:biasTruth} Bias in the systematics-corrected correlation function, relative to the sample standard deviation of the true correlation function. The orange dashed line shows the mean bias for the \num{100} uncontaminated mocks, and the orange shaded region is the error on the mean. Similarly, the blue solid line and shaded region are the mean and error on the mean for the \num{100} contaminated mocks. Note that there is a non-trivial bias even for the contaminated mocks indicating that we are over-correcting for SPs. We only show scales with $\theta > \SI{8}{\arcmin}$ for clarity. The gray shaded regions are once again the small scale cuts from \citetalias{2018PhRvD..98d2006E}.}
\end{figure*}

We seek to calibrate the amount of over-correction for our method based on the results from \cref{fig:corrections}. However, \emph{note that the level of over correction is itself sensitive to the input amount of contamination}. This is apparent in \cref{fig:biasTruth}, which shows the mean and error on the mean of the over-correction for both uncontaminated (orange) and contaminated (blue) mock galaxy catalogs.

We use the results in \cref{fig:biasTruth} to reduce the impact of over-correction, and to characterize the remaining systematic uncertainty associated with this effect. Because we see that the level of over-correction is sensitive to the amount of contamination and we do not know the actual contamination level in the data, we must account for this sensitivity when we de-bias. The contaminated and uncontaminated mocks represent the two extreme possibilities for the data, so we de-bias our correlation functions using the mean of the over-correction measured in the contaminated and uncontaminated mocks. That is, we define
%
\begin{linenomath}\begin{equation}
    \label{eq:bias}
    \Delta w(\theta) \equiv \frac{1}{2} \left[\langle w_{\rm corr0}^{\rm cont}(\theta) - w_{\rm true}(\theta)\rangle + \langle w_{\rm corr0}^{\rm uncont}(\theta) - w_{\rm true}(\theta)\rangle\right] ,
\end{equation}\end{linenomath}
%
where $w_{\rm corr0}^{\rm cont}(\theta)$ is the systematics-corrected correlation function at $\theta$ for the contaminated mock galaxy catalogs prior to de-biasing, and $w_{\rm corr0}^{\rm uncont}$ is the equivalent quantity computed for the uncontaminated mock galaxy catalogs. The average $\langle \cdot \rangle$ above is over the simulated data sets. The difference between the two terms in \cref{eq:bias} is indicative of the systematic uncertainty of this bias correction, as explained in \cref{sec:noise}. Given $\Delta w$, we define an updated systematics-corrected correlation function $w_{\rm corr1}$ via
%
\begin{linenomath}\begin{equation}
    \label{eq:corr1}
    w_{\rm corr1}(\theta) \equiv w_{\rm corr0}(\theta) - \Delta w(\theta)\, .
\end{equation}\end{linenomath}
%
The green dash-dotted line and shaded band in \cref{fig:corrections} show the mean and $1 \sigma$ region for the difference between our updated systematics-corrected correlation function estimates $w_{\rm corr1}$ and the true correlation function, as estimated using \num{100} contaminated mock catalogs. Recall that the y-axis is scaled in units of the purely statistical uncertainty of the DES Y1 analysis. It is clear from the figure that while a residual bias remains, the amplitude and uncertainty is much smaller than the statistical uncertainties for the DES Y1 data set. Moreover, the true underlying correlation function is within the expected errors in the measurement.

\section{The Impact of Systematics Removal on the Noise}
\label{sec:noise}

The covariance matrix used in \citetalias{2018PhRvD..98d2006E} when fitting the galaxy clustering signal was solely based upon theoretical considerations, as described in \citet*{2017arXiv170609359K}. In particular, it accounted only for Poisson noise and sample variance in the galaxy density field, where the latter includes both Gaussian and connected terms, as well as the super-sample covariance contribution. In practice, removing the imprint of systematic fluctuations on the galaxy density field carries with it additional uncertainty that needs to be propagated into the covariance matrix used to analyze the data. We now characterize this additional noise contribution.

We start with the statistical uncertainty in our method, the uncertainty due to the noise in our estimates of the linear coefficients of the SP maps. Because we use an MCMC to fit for the coefficients describing the impact of SPs, we can readily sample the posterior distribution to obtain realizations of the coefficients. For each such set of coefficients, we calculate the systematics-corrected correlation function $w_{\rm corr0}$, resulting in many realizations of systematics-corrected correlation functions. We calculate the covariance matrix from these realizations, and re-scale it by the factor of $\langle \chi^2\rangle/18=\lamfac$ from the discussion in \cref{sec:mockResults}. This defines the statistical covariance matrix $\mat{C^{\rm stat}}$, which characterizes statistical uncertainties in the systematics correction. We note that our estimation does not allow for statistical covariance in the SP corrections across the redshift bins. Since the noise in the coefficients of the SP maps depends on the galaxy density field, which will be correlated across bins, this is not true in detail. However, we expect the relative quality of the \redmagic{} photometric redshifts implies that any such correlation is small, particularly when propagated onto the SP coefficients.

\begin{figure}
    \centering
    \includegraphics[width=\linewidth,keepaspectratio]{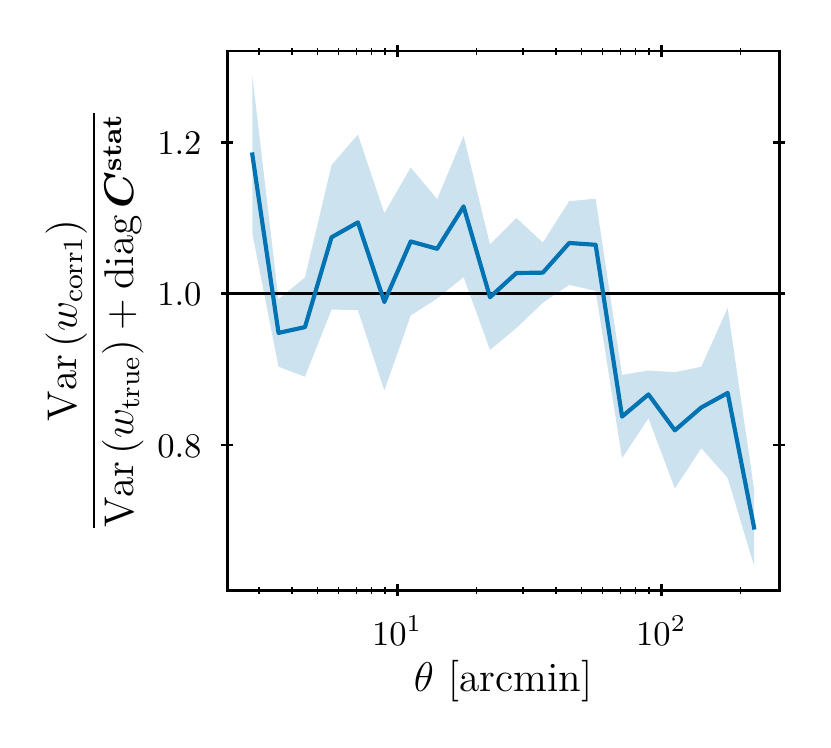}
    \caption{A comparison of $\var\left(w_{\mathrm{corr1}}\right)$ in our simulations and our revised error estimate $\var\left(w_{\mathrm{true}}\right) + \diag \mat{C^{\mathrm{stat}}}$. Our revised error estimate correctly describes the variance of the simulations at small scales, but slightly overestimates the noise at large scales. This likely reflects reduced sample variance due to removal of clustering modes at large scales.}
    \label{fig:cstat_debias_correction}
\end{figure}

In the process of de-biasing our systematics-corrected correlation functions, i.e. going from $w_{\mathrm{corr0}}$ to $w_{\mathrm{corr1}}$, our corrections may remove clustering modes from the correlation function. This will in turn remove some of the sample variance from $w_{\mathrm{corr1}}$, particularly at large scales. To test for this possibility, we compared the variance in $w_{\mathrm{corr1}}$ as measured in our simulations to the quantity
\begin{equation*}
    \hat{\var}\left(w_{\mathrm{corr1}}\right) \equiv \var\left(w_{\mathrm{true}}\right) + \diag \mat{C^{\mathrm{stat}}} ,
\end{equation*}
where $\mat{C^{\mathrm{stat}}}$ is the statistical covariance matrix including the re-scaling from the previous paragraph. The blue shaded region in \cref{fig:cstat_debias_correction} shows the ratio of $\var\left(w_{\mathrm{true}}\right) + \diag \mat{C^{\mathrm{stat}}}$ and $\hat{\var}\left(w_{\mathrm{corr1}}\right)$ averaged over all redshift bins. The width of the band represents the $1-\sigma$ region around the mean. We see that at small scales the variance in our simulations is consistent with our error estimate. By contrast, we overestimate the error at large scales, likely due to the removal of clustering modes in our systematic correction algorithm. We have found that accounting for this reduced variance tends to make the covariance matrix of the \tct{} data vector non-invertible due to the cross terms between probes. Because the systematic bias in the variance estimate is small, and because our revised errors have little impact on cosmological posteriors (see below), we will leave the problem of how to adequately model the resulting decrease in variance to future work.

The systematic uncertainty associated with our de-biasing procedure of \cref{sec:calibration} is calculated as the sum in quadrature of two distinct terms. The first term sets the systematic uncertainty to half the amplitude of the applied correction, i.e. large corrections will result in large uncertainties. The second term accounts for the difference in the amount of over-correction inferred from the contaminated and uncontaminated mocks. If the inferred over-corrections are vastly different, the resulting mean correction should be assigned a large uncertainty. This uncertainty is set to half the difference between the over-correction inferred from the contaminated and uncontaminated mocks. The corresponding covariance matrix characterizing these systematic uncertainties takes the form
%
\begin{linenomath}\begin{equation}
    \label{eq:sysCov}
    \mat{C^{\rm sys}}_{ab} \equiv \frac{1}{4} \left[\Delta w(\theta_a) \Delta w(\theta_b) + \delta w(\theta_a) \delta w(\theta_b)\right]\, ,
\end{equation}\end{linenomath}
%
where $a$ and $b$ index angular bins, and where we have defined
%
\begin{linenomath}\begin{equation*}
    \delta w(\theta) \equiv \frac{1}{2} \left[ \left\langle w_{\rm corr0}^{\rm cont}(\theta) - w_{\rm true}(\theta)\right\rangle - \left\langle w_{\rm corr0}^{\rm uncont}(\theta) - w_{\rm true}(\theta) \right\rangle\right]\, .
\end{equation*}\end{linenomath}
%
As in \cref{eq:bias}, the average $\langle \cdot \rangle$ above is over all simulated data sets.

\begin{figure*}
    \centering
    \includegraphics[width=\textwidth,keepaspectratio]{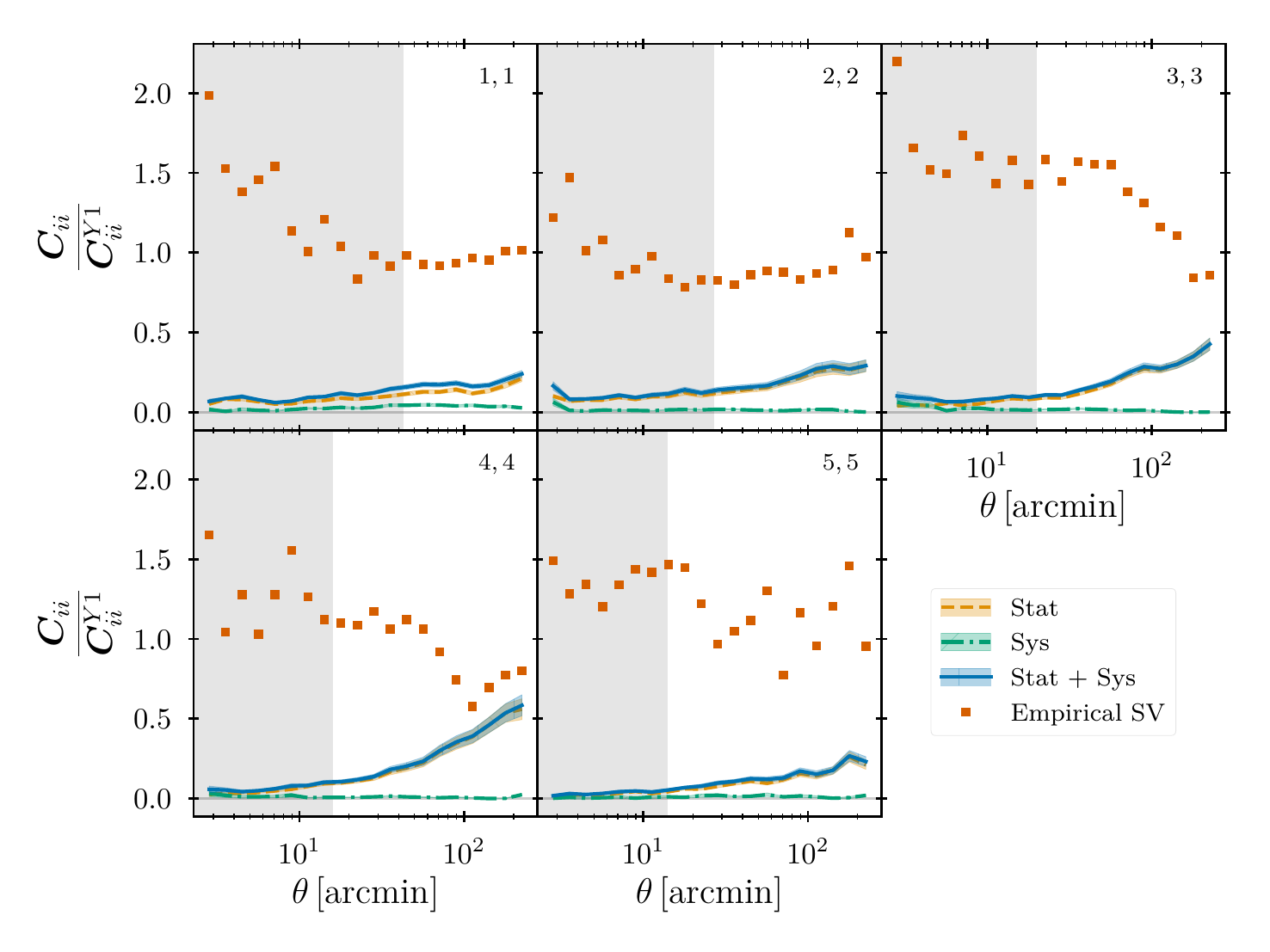}
    \caption{\label{fig:covComponents} A comparison of the diagonal elements from various components of the covariance matrix relative to the diagonal elements of the theoretical covariance matrix utilized in \citetalias{2018PhRvD..98d2006E}. In all cases, the denominator of the quantity on the y-axis is the diagonal elements of $\mat{C^{Y1}}$. The numerator for the blue solid line corresponds to our updated error estimates, including both statistical (orange dashed) and systematic (green dash-dotted) errors. The red squares show the corresponding empirical estimates of the variance in our simulations. The gray shaded region is the small scale cut from \citetalias{2018PhRvD..98d2006E}, and will not impact the cosmology results.}
\end{figure*}

The final covariance matrix estimate for the data is $\mat{C^{Y1}} + \mat{C^{\rm stat}} + \mat{C^{\rm sys}}$, where $\mat{C^{Y1}}$ is the theoretical covariance matrix used in \citetalias{2018PhRvD..98d2006E}. The green dash-dotted line and band in \cref{fig:covComponents} show the mean and uncertainty of the ratio between the diagonal elements of $\mat{C^{\rm sys}}$, as defined in \cref{eq:sysCov}, to the diagonal elements of $\mat{C^{Y1}}$. The orange dashed line and band is the same ratio but for $\mat{C^{\rm stat}}$. We have checked that increasing the number of realizations used to estimate $\mat{C^{\rm stat}}$ does not significantly change our measured covariance. The combination of the systematic and statistical covariance relative to the Y1 covariance is shown as the blue solid line and band. The gray shaded region in each panel shows the region excluded by the small scale cuts for the cosmology analysis in \citetalias{2018PhRvD..98d2006E}, for which our changes will not impact the inferred cosmological parameters. While uncertainties in our de-biasing procedure for over-correction are negligible, we see that the statistical uncertainties in our systematics mitigation algorithm start to become comparable to statistical uncertainties in the correlation function at large scales.

\section{Results}
\label{sec:results}

As a brief summary of \cref{sec:method,sec:mocks,,sec:noise}, we assume fluctuations in SPs introduce artificial galaxy fluctuations through a local linear response. We calibrate these response coefficients using the observed galaxy density maps and SP maps, and use them to remove the impact of systematic fluctuations in the galaxy density field. Using mock galaxy catalogs, we demonstrate that our method results in some small amount of over-correction, which we calibrate. We further characterize the additional statistical and systematic uncertainty introduced by our systematics-mitigation algorithm. We now apply our full systematics-correction algorithm to the DES Y1 data set.

\begin{figure*}
    \centering
    \includegraphics[width=\textwidth,height=0.5\textheight,keepaspectratio]{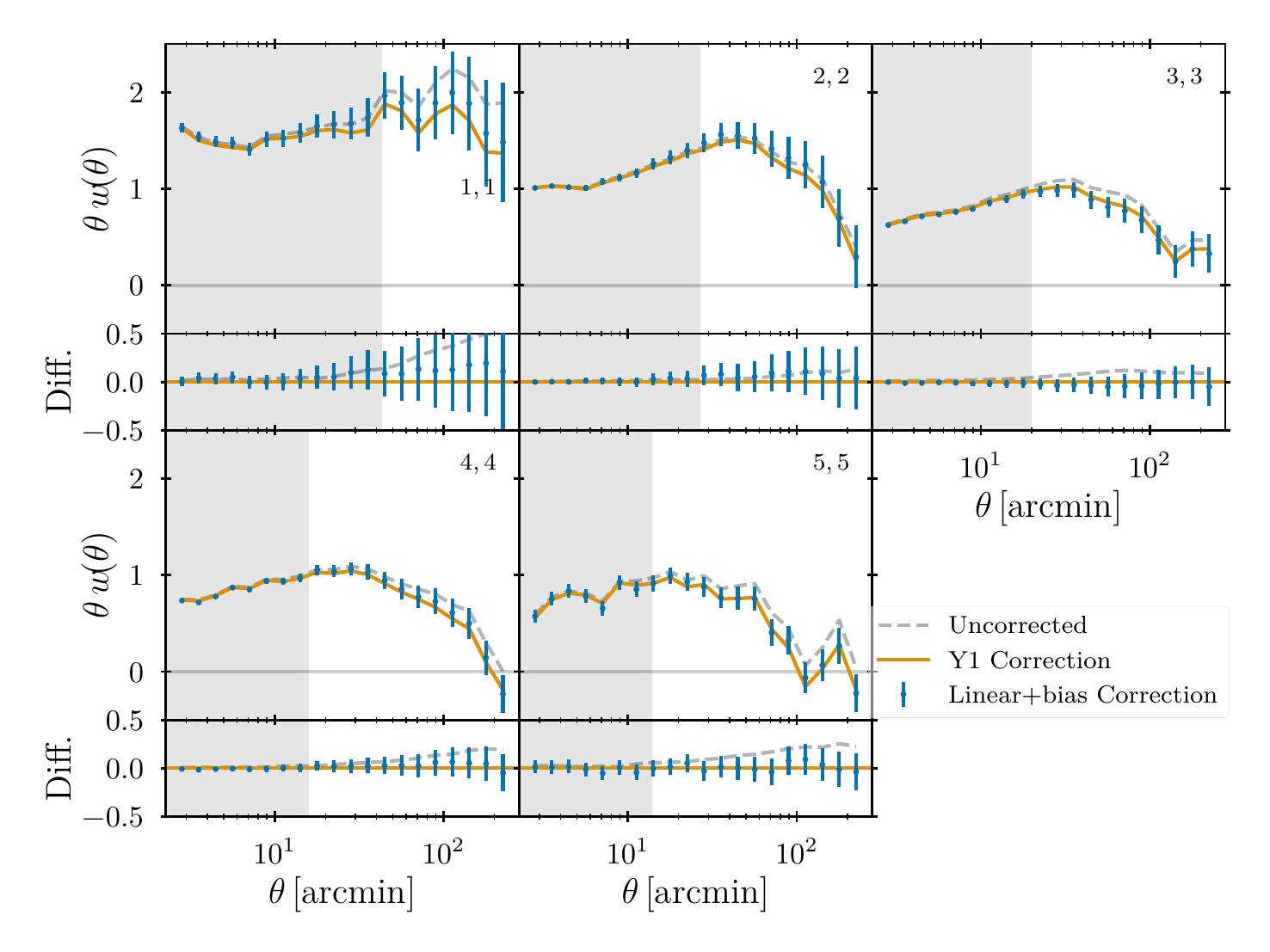}
    \caption{\label{fig:wthetaY1} The correlation function in each redshift bin for the DES Y1 \redmagic{} galaxies. The gray dashed line is the correlation function without correcting for SPs. The orange solid line is the systematics-corrected correlation from \citetalias{2018PhRvD..98d2006E}. The blue points are the de-biased correlation function using our linear model weights, and the error bars are obtained from the full ($\mat{C^{Y1}} + \mat{C^{\rm stat}} + \mat{C^{\rm sys}}$) covariance matrix. Note that while the gray and orange lines are computed with the DES Y1 mask, the blue points use our restricted mask with $\delta_{\rm sys} \leq 0.2$, resulting in \SI{\sim 3.5}{\percent} less area.}
\end{figure*}

In \cref{fig:wthetaY1}, we show the angular correlation function in each of the five redshift bins using our systematics weights and bias correction as blue circles, with errors from the combined $\mat{C^{Y1}} + \mat{C^{\rm stat}} + \mat{C^{\rm sys}}$ covariance matrix. For comparison, we also show the correlation function without correction  and the systematics-corrected correlation function from \citetalias{2018PhRvD..98d2006E}. We note that in arriving at our updated correlation function, there is a small change in the mask to mitigate the impact of non-linear systematic fluctuations, so that the areas over which the correlation functions are computed are not precisely the same. The bottom panel in each figure shows the difference of each of the correlation function relative to the systematics-corrected estimate of \citetalias{2018PhRvD..98d2006E}. We see that the two different methods for estimating systematic corrections are in excellent agreement relative to the statistical uncertainty of the DES Y1 data set. Nevertheless, some small differences are clearly present. It is interesting to note that in the second redshift bin, our correction results in slightly \emph{more} correlation than the uncorrected correlation function, rather than \emph{less}. This boost is due to the over-correction de-biasing procedure calibrated in the mocks.

\begin{table}
    \centering
    \begin{tabular}{lccccc}
        \toprule
        $z$ bin & $\chi^2_{\rm stat+sys}$ & $\chi^2_{\rm tot}$ & $\chi^{\prime 2}_{\rm stat+sys}$ & $\chi^{\prime 2}_{\rm tot}$ & \# Bins \\ \midrule
        1 & 18.02(0.02) & 0.463 & 9.563(0.30) & 0.286 & 8 \\
        2 & 120.0(0.00) & 2.04 & 57.03(0.00) & 1.37 & 10 \\
        3 & 97.46(0.00) & 0.900 & 26.71(0.01) & 0.502 & 11 \\
        4 & 45.65(0.00) & 0.922 & 67.91(0.00) & 0.611 & 12 \\
        5 & 344.6(0.00) & 1.89 & 22.82(0.04) & 0.354 & 13 \\ \bottomrule
    \end{tabular}
    \caption{\label{tab:wChi2} The $\chi^2$ for the systematics-corrected correlation function from \citetalias{2018PhRvD..98d2006E} and this work in each redshift bin. The last column is the number of angular bins used to calculate the $\chi^2$, which are the bins outside the small scale cut represented by the gray shaded regions in \cref{fig:wthetaY1}. The second column is the $\chi^2$ when including only the uncertainty from the systematics correction, while the third column is the $\chi^2$ relative to the full covariance matrix. The fourth and fifth columns are the same as the second and third, but the $\delta_{\rm sys} > 0.2$ mask is applied to the galaxies with the DES Y1 weights. The numbers in parentheses in the second and fourth columns show the probability to exceed the $\chi^2$ given the number of angular bins in the last column (the probability to exceed is \num{\sim 1.0} for all bins in both the third and fifth columns). Our updated correlation function is only consistent with that from \citetalias{2018PhRvD..98d2006E} in the first redshift bin, but it is consistent with the correlation function with our mask and the Y1 weights in bins \numlist{1;3;5}.}
\end{table}

To quantify the difference in the correlation functions from the two different weighting methods, \cref{tab:wChi2} shows the $\chi^2$ statistic for the DES Y1 correlation function and our correlation function, namely 
%
\begin{linenomath}\begin{equation*}
    \chi^2 = \left(w_{Y1}(\theta) - w_{\rm corr1}(\theta)\right)^\transpose \mat{C}^{-1} \left(w_{Y1}(\theta) - w_{\rm corr1}(\theta)\right),
\end{equation*}\end{linenomath}
%
where the choice of covariance matrix $\mat{C}$ used requires some discussion (see below). In calculating $\chi^2$, we exclude any angular bins that are removed with the small scale cut (the gray regions in \cref{fig:wthetaY1}). The number of remaining angular bins after the small scale cut is shown in the last column of the table. The difference between the correlation functions should not be subject to Poisson noise or sample variance, as these are the same for both correlation functions. Therefore, in the second column of \cref{tab:wChi2}, we show the $\chi^2$ when we use $\mat{C} = \mat{C^{\rm stat}} + \mat{C^{\rm sys}}$, with the probability to exceed the given $\chi^2$ shown in parentheses. While in principle this comparison should also be subject to the uncertainty due to the method of \citetalias{2018PhRvD..98d2006E}, that paper demonstrated that the uncertainties in their systematics correction didn't impact the cosmological priors and therefore those uncertainties were not characterized. Consequently, our comparison does not account for the uncertainty in the Y1 systematics correction. It is clear that our weights method results in a correlation function that is formally inconsistent with that of the Y1 analysis assuming zero uncertainty from the Y1 weights method. However, the size of the cosmology contours is sensitive to the full covariance matrix $\mat{C^{Y1}} + \mat{C^{\rm stat}} + \mat{C^{\rm sys}}$. The third column in \cref{tab:wChi2} shows the $\chi^2$ when we use the full covariance matrix for $\mat{C}$. Notice that in this case, the $\chi^2/\mathrm{dof} \leq 0.1$ for most redshift bins. This result explicitly demonstrates that the difference in the correlation function produced by the two  methods is small relative to the statistical uncertainty.

While our updated correlation function is inconsistent with the correlation function presented in \citetalias{2018PhRvD..98d2006E} when excluding the statistical uncertainty, the difference between them is actually sourced by two effects: the difference in the weights produced by the two corrections, and the difference in the footprint. In particular, removal of the pixels with $\delta_{\rm sys}^i > 0.2$ improves the agreement between the two. To determine whether the corrections are consistent after accounting for the different masks, we recompute the galaxy correlation function using the fiducial Y1 weights over our new mask. The $\chi^2$ statistics for this comparison are shown in the fourth and fifth columns of \cref{tab:wChi2}. As before, the $\chi^2_{\rm tot}/\mathrm{dof} \lesssim 0.1$ in all bins when using the full covariance matrix. However, the correlation functions with the different weights in this case are in much better agreement. We take the two systematic corrections to be consistent with one another if the probability to exceed the observed $\chi^2$ between them is at least \num{0.01}. Based on this definition, the corrections for redshift bins \numlist{1;3;5} are consistent with each other, as opposed to only the first redshift bin when the masks were different. The second and fourth bins are inconsistent in both cases.

\begin{figure}
    \centering
    \includegraphics[width=\columnwidth,keepaspectratio]{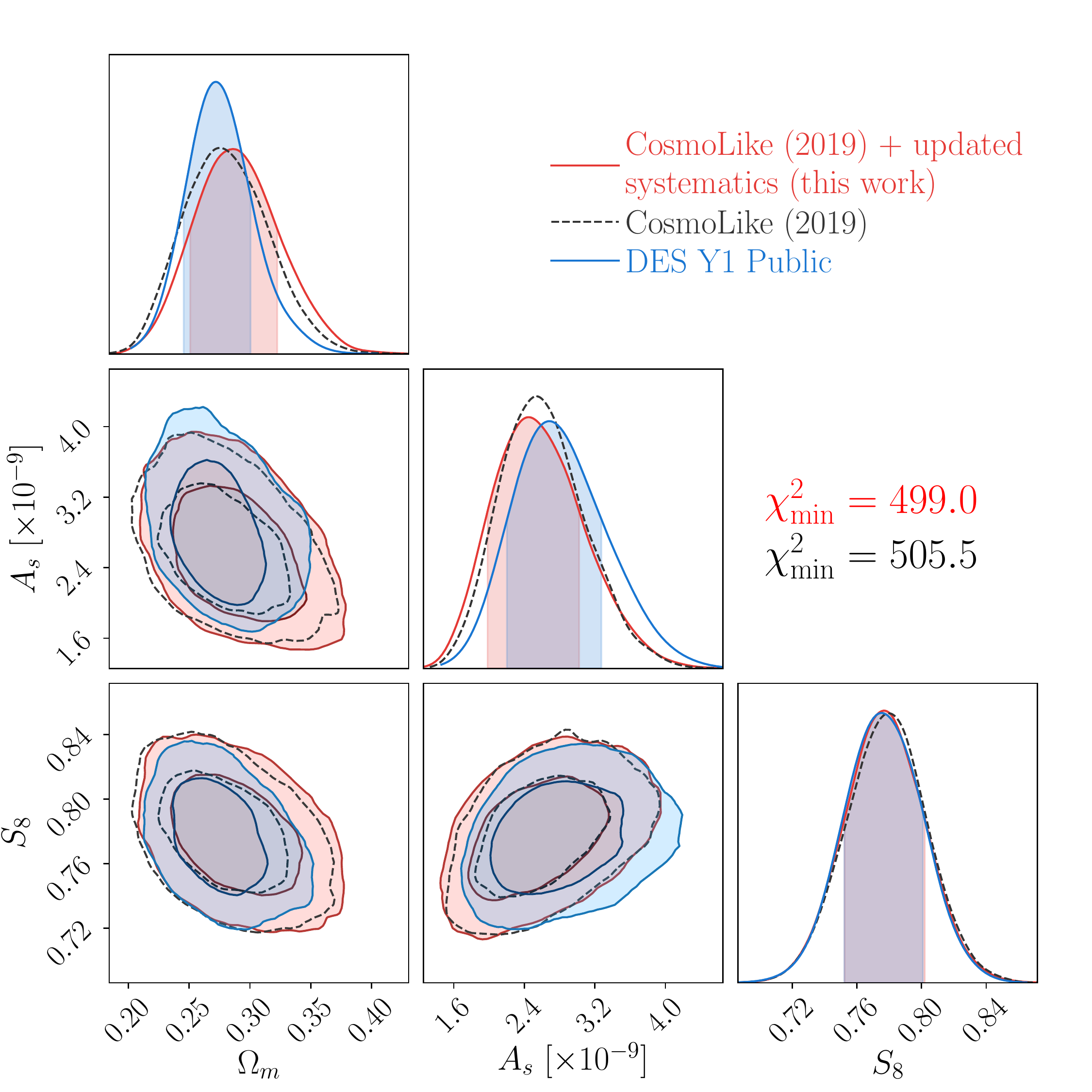}
    \caption{\label{fig:3x2ptContours} A comparison of the cosmology contours for the \tct{} analysis, with each 2-dimensional contour showing the \SI{68}{\percent} and \SI{95}{\percent} confidence levels, and the shaded regions in the 1-dimensional plots signifying the \SI{68}{\percent} confidence level. The blue contours are the public DES Y1 results as in \citet{2018PhRvD..98d3526A}. The red contours are the results with our new correlation function and updated covariance matrix. Note that the blue and red contours use a different version of \texttt{CosmoLike} and different samplers. The black dashed lines also show the contours using the DES Y1 data vector, but using the same version of \texttt{CosmoLike} and same sampler as was used to generate the red contours. The minimum $\chi^2$ for the DES Y1 data vector and our updated data vector are shown as the black and red text, respectively, with \num{444} degrees of freedom.}
\end{figure}

We use our new de-biased systematics-corrected correlation function and full $\mat{C^{Y1}} + \mat{C^{\rm stat}} + \mat{C^{\rm sys}}$ covariance matrix in combination with the cosmic shear and galaxy-galaxy lensing data vectors and covariance matrices from the DES Y1 cosmology analysis \citep{2018PhRvD..98d3526A} to re-run the DES \tct{} cosmology analysis. The resulting cosmology contours for $\Omega_m$, $A_s$, and $S_8$ are shown in blue in \cref{fig:3x2ptContours}. For our analysis, we use an updated version of \texttt{CosmoLike} \citep{2017MNRAS.470.2100K,2020MNRAS.497.2699F} and use \texttt{emcee} \citep{2013PASP..125..306F} as our sampler. The result with this pipeline and our updated data vector and covariance matrix are shown in red in \cref{fig:3x2ptContours}.

As we use a different pipeline and sampler than the fiducial Y1 analysis of \citet{2018PhRvD..98d3526A}, it is unclear how much of the difference between the red and blue contours in \cref{fig:3x2ptContours} is because of our changes to the data vector and covariance matrix and how much is a reflection of the differences in the modelling pipeline. We therefore show as black dashed lines in \cref{fig:3x2ptContours} the results of using the updated \texttt{CosmoLike} pipeline when run on the fiducial Y1 data vector and covariance matrix. The differences between the red and black contours are due to the difference in the estimated correlation function and its corresponding covariance matrix. It is clear that our weighting method does not have a significant impact on the cosmological inference relative to the Y1 analysis. This is expected given that both the difference in the correlation functions with the two different weighting methods and the uncertainty in our systematic correction are small relative to the statistical uncertainty of the measurement.

The black and red text above the histogram of $S_8$ in \cref{fig:3x2ptContours} show the minimum $\chi^2$ values for the fiducial Y1 data vector and our updated data vector, respectively, for each data vector compared to the model with \num{444} degrees of freedom \citep[see][]{2018PhRvD..98d3526A}. The minimum $\chi^2$ in each case is $-2 \log L_{\rm max}$ at the maximum likelihood point in the MCMC chain. It is encouraging to see that even though our method does not significantly change the cosmological inference, it does result in a significant improvement in the goodness of fit ($\Delta \chi^2 = -6.5$ with no additional parameters). This improvement in the $\chi^2$ is due to both the increased error from our systematics correction and the shifts in the data vector that occur when replacing the Y1 weighting method with ours. To show that this is the case, we consider the calculation of the best fit $\chi^2$ with our updated data vector and covariance matrix, which we now write as
%
\begin{linenomath}\begin{equation*}
    \chi^2_{\rm new} = \left(\vvec{d}_{Y1} + \vvec{\Delta} - \vvec{m}_{Y1}\right)^\transpose \left(\mat{C^{Y1}} + \mat{\delta C}\right)^{-1} \left(\vvec{d}_{Y1} + \vvec{\Delta} - \vvec{m}_{Y1}\right),
\end{equation*}\end{linenomath}
%
where $\vvec{d}_{Y1}$ is the original data vector from the Y1 analysis, $\vvec{m}_{Y1}$ is the best fit model vector from the original Y1 analysis, $\mat{\delta C} \equiv \mat{C^{\rm stat}} + \mat{C^{\rm sys}}$ is the change in the covariance matrix, and $\vvec{\Delta}$ is the change to the difference between the data vector and best fit model vector introduced by our weights method. Note that this means that $\vvec{\Delta}$ is sensitive to both the change in the data vector as well as changes to the best fit parameters. We can expand this equation around $\mat{\delta C} = 0$, dropping terms that are beyond first order in $\mat{\delta C}$ as well as terms involving $\vvec{\Delta}^\transpose \mat{\delta C}$. Doing so, we find
%
\begin{linenomath}\begin{align*}
    \Delta \chi^2 \approx &\, \vvec{\Delta}^\transpose \left(\mat{C^{Y1}}\right)^{-1} \left[2 \left(\vvec{d}_{Y1} - \vvec{m}_{Y1}\right) + \vvec{\Delta}\right] \\ - &\left[\left(\mat{C^{Y1}}\right)^{-1} \left(\vvec{d}_{Y1} - \vvec{m}_{Y1}\right)\right]^\transpose \mat{\delta C} \left[\left(\mat{C^{Y1}}\right)^{-1} \left(\vvec{d}_{Y1} - \vvec{m}_{Y1}\right)\right].
\end{align*}\end{linenomath}
%
The first term in this expression gives the $\Delta \chi^2$ resulting from changing the data vector and the difference in the resulting best fit model vector. The second term is the $\Delta \chi^2$ caused by the change to the covariance matrix from our systematics correction. We find $\Delta \chi^2 \approx -3.6$ for the first term and $\Delta \chi^2 \approx -3.2$ for the second. From this, we conclude that both the shift in the data vector (and resulting shift in the best fit) and the increased uncertainty due to our systematics correction contribute to the improvement in the fit.

\section{Conclusions}
\label{sec:conclusions}

We have presented a method for using a linear model to mitigate the effect of systematic fluctuations in galaxy clustering analyses due to observing conditions. Our method uses a Gaussian likelihood to fit the linear model to the observed galaxy over-density and SP maps on each pixel. Our analysis explicitly incorporates the fact that neighboring pixels in the sky are correlated using an iterative approach: our first iteration uses a diagonal covariance matrix, while the second builds a non-diagonal covariance matrix from the systematics-corrected correlation function estimated from the first iteration. We further use mock catalogs to calibrate the remaining over-correction bias, which we then remove from the data correlation function.

We apply our methodology to the DES Y1 \redmagic{} data set. Our method has four important advantages relative to that adopted in the DES \tct{} analysis presented in \citet{2018PhRvD..98d3526A}, namely:
\begin{itemize}
    \item Our method does not require that decisions be made with regards to which survey properties matter and which don't. This also allows for the possibility of multiple survey properties ``conspiring'' to create an observationally significant signal without any single systematic reaching that threshold.
    \item Our method properly increases the error budget of the galaxy correlation function estimate by accounting for the statistical and systematic uncertainty associated with systematic mitigation. We have found doing so non-trivially impacts the goodness-of-fit statistic of the best fit cosmological model.
    \item Our method explicitly incorporates clustering information from neighboring pixels in our calibration of the impact of survey properties on the galaxy density field.
    \item Our method is fully automated: it can be run from start to finish with minimal supervision, enabling for quick turn around for future data sets, with no extra tuning.
\end{itemize}

While our updated systematics-corrected correlation function in the DES Y1 data set is formally inconsistent with that of \citetalias{2018PhRvD..98d2006E}, the two are in good agreement relative to the level of statistical uncertainty in DES Y1, and the corrections are consistent in redshift bins \numlist{1;3;5} when applied over the same footprint. Because the statistical uncertainty in the measurement is larger than the uncertainty in our correction, we observe no significant impact on the cosmological inference using a data vector with our systematics weights relative to the Y1 \tct{} cosmology analysis. Encouragingly, however, we do see an improvement in the goodness of fit, which is caused by both the change to the data vector with our new weights and the increased error from our systematics correction. We also expect the difference in the data vector and the uncertainty in the correction to become more important in the near future as the large number of galaxies observed by upcoming surveys decreases the statistical uncertainty in galaxy clustering measurements.

\section*{Acknowledgements}

ELW and ER were supported by the DOE grant DE-SC0015975. XF is supported by NASA ROSES ATP 16-ATP16-0084 grant. Some calculations in this paper use High Performance Computing (HPC) resources supported by the University of Arizona TRIF, UITS, and RDI and maintained by the UA Research Technologies department.

This paper has gone through internal review by the DES collaboration.

Funding for the DES Projects has been provided by the U.S. Department of Energy, the U.S. National Science Foundation, the Ministry of Science and Education of Spain, 
the Science and Technology Facilities Council of the United Kingdom, the Higher Education Funding Council for England, the National Center for Supercomputing 
Applications at the University of Illinois at Urbana-Champaign, the Kavli Institute of Cosmological Physics at the University of Chicago, 
the Center for Cosmology and Astro-Particle Physics at the Ohio State University,
the Mitchell Institute for Fundamental Physics and Astronomy at Texas A\&M University, Financiadora de Estudos e Projetos, 
Funda{\c c}{\~a}o Carlos Chagas Filho de Amparo {\`a} Pesquisa do Estado do Rio de Janeiro, Conselho Nacional de Desenvolvimento Cient{\'i}fico e Tecnol{\'o}gico and 
the Minist{\'e}rio da Ci{\^e}ncia, Tecnologia e Inova{\c c}{\~a}o, the Deutsche Forschungsgemeinschaft and the Collaborating Institutions in the Dark Energy Survey. 

The Collaborating Institutions are Argonne National Laboratory, the University of California at Santa Cruz, the University of Cambridge, Centro de Investigaciones Energ{\'e}ticas, 
Medioambientales y Tecnol{\'o}gicas-Madrid, the University of Chicago, University College London, the DES-Brazil Consortium, the University of Edinburgh, 
the Eidgen{\"o}ssische Technische Hochschule (ETH) Z{\"u}rich, 
Fermi National Accelerator Laboratory, the University of Illinois at Urbana-Champaign, the Institut de Ci{\`e}ncies de l'Espai (IEEC/CSIC), 
the Institut de F{\'i}sica d'Altes Energies, Lawrence Berkeley National Laboratory, the Ludwig-Maximilians Universit{\"a}t M{\"u}nchen and the associated Excellence Cluster Universe, 
the University of Michigan, NFS's NOIRLab, the University of Nottingham, The Ohio State University, the University of Pennsylvania, the University of Portsmouth, 
SLAC National Accelerator Laboratory, Stanford University, the University of Sussex, Texas A\&M University, and the OzDES Membership Consortium.

Based in part on observations at Cerro Tololo Inter-American Observatory at NSF's NOIRLab (NOIRLab Prop. ID 2012B-0001; PI: J. Frieman), which is managed by the Association of Universities for Research in Astronomy (AURA) under a cooperative agreement with the National Science Foundation.

The DES data management system is supported by the National Science Foundation under Grant Numbers AST-1138766 and AST-1536171.
The DES participants from Spanish institutions are partially supported by MICINN under grants ESP2017-89838, PGC2018-094773, PGC2018-102021, SEV-2016-0588, SEV-2016-0597, and MDM-2015-0509, some of which include ERDF funds from the European Union. IFAE is partially funded by the CERCA program of the Generalitat de Catalunya.
Research leading to these results has received funding from the European Research
Council under the European Union's Seventh Framework Program (FP7/2007-2013) including ERC grant agreements 240672, 291329, and 306478.
We  acknowledge support from the Brazilian Instituto Nacional de Ci\^encia
e Tecnologia (INCT) e-Universe (CNPq grant 465376/2014-2).

This manuscript has been authored by Fermi Research Alliance, LLC under Contract No. DE-AC02-07CH11359 with the U.S. Department of Energy, Office of Science, Office of High Energy Physics.

Some of the results in this paper have been derived using the \texttt{healpy} and \texttt{HEALPix} packages. This research made use of Astropy,\footnote{http://www.astropy.org} a community-developed core Python package for Astronomy \citep{2013A&A...558A..33A,2018AJ....156..123A}.

\section*{Data Availability} 
The DES Y1 \redmagic{} catalog is available for download at \url{https://des.ncsa.illinois.edu/releases/y1a1/key-catalogs/key-redmagic}. The DES observing condition maps, excluding dust and stellar density, are available for download at \url{https://des.ncsa.illinois.edu/releases/y1a1/gold/systematics}. The dust map is available from Planck at \url{https://irsa.ipac.caltech.edu/data/Planck/release_1/all-sky-maps/previews/HFI_CompMap_ThermalDustModel_2048_R1.20/index.html}. The stellar density catalog is not publicly available but can be constructed from the DES Y1GOLD badmask (\url{https://des.ncsa.illinois.edu/releases/y1a1/gold/footprint}) and the first public data release (\url{https://des.ncsa.illinois.edu/releases/dr1}). It may also be made available upon request with permission of the Dark Energy Survey Collaboration.



\bibliographystyle{mnras_2author}
\bibliography{references}








\bsp	
\label{lastpage}
\end{document}